\documentclass[1p,twoside,number,sort&compress,floatfix,final]{elsarticle}

\usepackage{amsmath,amssymb}
\usepackage{stix,bm}

\usepackage[colorlinks]{hyperref}
\newcommand{\ie}{i.{\kern1pt}e.}
\newcommand{\aka}{a.{\kern1pt}k.{\kern1pt}a.}
\newcommand{\iid}{i.{\kern1pt}i.{\kern1pt}d.}

\newcommand{\mku}{\mkern1mu}

\newcommand{\eps}{\ensuremath{\varepsilon}}
\newcommand{\abs}[1]{\left|#1\right|}

\DeclareMathOperator{\PP}{\mathbb{P}}
\DeclareMathOperator{\EE}{\mathbb{E}}
\DeclareMathOperator{\var}{Var}
\DeclareMathOperator{\RSS}{RSS}

\begin{document}

\begin{frontmatter}

\title{Longest weakly increasing subsequences of discrete random walks on the integers with heavy tailed distribution of increments}

\author{José Ricardo G. Mendonça}
\ead{jricardo@usp.br}

\author{Marcelo V. Freire}
\ead{mvf@usp.br}

\address{Escola de Artes, Ciências e Humanidades, Universidade de São Paulo \\ Rua Arlindo Bettio 1000, Vila Guaraciaba, 03828-000 São Paulo, SP, Brazil}

\begin{abstract}
We investigate the behavior of the length of the longest weakly increasing subsequences (weak LIS) of $n$-step random walks with nonzero integer increments $k = \pm 1, \pm 2, \dots$ given by a symmetric heavy tailed mass distribution proportional to $\abs{k}^{-1-\alpha}$ for several values of the real parameter $\alpha > 0$ together with that of the simple random walk ($k=\pm 1$), to which the $n$-step heavy tailed walks reduce when $\alpha$ grows large enough that step jumps beyond $\pm 1$ become essentially absent on the scale of $n$.
By means of exploratory fits, weighted nonlinear least squares, and nested-model comparisons, we found that the sample average length $\langle{L_{n}}\rangle$ scales like $\langle{L_{n}}\rangle \sim \sqrt{n}\mku\log{n}$ when the distribution of increments has finite variance ($\alpha > 2$) and $\langle{L_{n}}\rangle \sim n^{\theta}$ with a varying exponent $\theta > 0.5$ when the variance is infinite ($\alpha \leq 2$). Distributional diagnostics indicate that the bulk of the $L_{n}$ distribution is very well-approximated by a lognormal model, though systematic deviations are observed in the tails.
Our results corroborate and expand upon previous results for the LIS of other types of heavy-tailed random walks and raise a conjecture as to whether the distribution of $L_{n}$ is given, or can be effectively described, by a lognormal distribution.
\end{abstract}


\end{frontmatter}


\section{\label{intro}Introduction}

The investigation of the longest increasing subsequences (LIS) of random permutations (\aka\ Ulam's problem) \cite{ulam,hammersley} has spurred the development of a host of techniques in group theory, combinatorics, statistics, and probability \cite{aldous,bdj99,patience,johansson,romik}, which, in turn, helped elucidate several issues in modern nonequilibrium statistical mechanics \cite{krug,corwin,takeuchi}. The LIS of random walks and other correlated time series, however, have remained little explored until recently \cite{angel,pemantle}, despite their potential applications in fields like statistical tests of independence and data stream analytics \cite{jesus2014,liben,gopalan2010,bonomi2016}. 

In a couple of recent papers, several properties of the LIS of random walks---their scaling behavior for different distributions of increments (short, heavy, and ultra heavy tailed), some universal properties of their distribution functions, large deviations, and abundance---have been investigated and some numerical conjectures have been raised \cite{lisjpa,hartmann,lispre,ultrafat}. Most of these studies, however, deal with continuous distributions of increments. It is expected, based on rigorous arguments and numerical evidence, that the scaling behavior of the LIS of discrete and continuous random walks differ. Moreover, for discrete random walks the longest weakly increasing and strictly increasing subsequences display different behavior \cite{angel}. It thus appears worthwhile to study the LIS of discrete random walks in general as well as the LIS of discrete random walks with heavy tailed distribution of increments, which have not yet received attention in the literature.

In this paper we numerically investigate the behavior of the length $L_{n}$ of the longest weakly increasing subsequence (the weak LIS) of symmetric heavy tailed random walks on the integers as a function of the walk length~$n$ and the tail index $\alpha > 0$ of a Pareto-like (or Zipf) distribution of increments. Using weighted nonlinear least squares and analysis of variance (ANOVA), we complement the exploratory fits of \cite{lisjpa,hartmann,lispre,ultrafat} and are able to objectively discriminate between the competing scaling forms $\sqrt{n}\mku\log{n}$ and $n^{\theta}$, an issue left unresolved in earlier work. The paper is organized as follows. In Section~\ref{sec:rw} we define the discrete random walks, their weak LIS, and the heavy tailed distributions that we employ in our study together with a numerical procedure to generate the appropriate random deviates, and in Section~\ref{sec:facts} we briefly review a few known facts about the LIS of random walks. In Section~\ref{sec:lis} we describe and discuss our numerical results by means of an exploratory analysis, and in Section~\ref{sec:anova} we complement the analysis with a weighted nonlinear least squares fit and formal hypothesis tests. In Section~\ref{sec:dist} we examine the distribution of $L_{n}$ itself, an issue that has been only tentatively addressed in the literature so far. Finally, in Section~\ref{sec:discuss} we summarize and discuss our results. 

A reader looking for the essential results without the full methodological apparatus will find them in Figure~\ref{fig:lis} (the object of study), Figure~\ref{fig:theta} (the position of this work in the broader landscape of LIS exponents), Table~\ref{tab:summary} (a one-glance summary of which scaling form holds in each regime), and Section~\ref{sec:discuss} (the qualitative conclusions and open questions).


\section{\label{sec:rw}The discrete heavy tailed random walk on the integers}

An $n$-step random walk on the integers is a sequence of random variables $S_{1}, \dots, S_{n}$ given by
\begin{equation}
\label{eq:rw}
S_{0}=0,\ S_{i} = S_{i-1}+X_{i},\ 1 \leq i \leq n,
\end{equation}
where the step increments $X_{i}$ are independent integer-valued random variables identically distributed according to some symmetric probability mass function $\PP(X=k) = \phi_{\alpha}(k)$, where $k$ is an integer and $\alpha$ stands for the parameter of the distribution. The $S_{i}$ are correlated random variables, since $\EE(S_{i}S_{j}) = \min(i,j)\EE(X^{2})$. 

Given a random walk $S_{1}, \dots, S_{n}$, its weak LIS is the longest possible weakly increasing subsequence $S_{i_{1}} \leq \cdots \leq S_{i_{L}}$ with $1 \leq i_{1} < \cdots < i_{L} \leq n$. There can be more than one weak LIS for any given sequence of numbers, with different elements but the same maximal length $L$. The LIS of a sequence of~$n$ numbers can be computed by the patience sorting algorithm, which employs dynamic programming and binary search, in $O(n)$ space and $O(n\log{n})$ time \cite{patience,fredman,sergei}.

We are interested in the behavior of the weak LIS of the random walk (\ref{eq:rw}) with a heavy-tailed distribution of increments, defined by an algebraic decay of $\phi_{\alpha}(k)$ for large $\abs{k}$. We can generate such random variables by the method of inversion by truncation of a continuous random variable \cite[Sec.~III.2]{devroye}. If $U$ is a uniform random variable in $(0,1)$, then for any given real number $\alpha>0$ the random variable $X = \lfloor U^{-1/\alpha} \rfloor$, where the floor $\lfloor x \rfloor$ evaluates to the greatest integer less than or equal to $x$, is distributed over the positive integers $k \geq 1$ with probability masses 
\begin{equation}
\label{eq:prb}
\phi_{\alpha}(k) = \PP(k \leq U^{-1/\alpha} < k+1) = 
\PP((k+1)^{-\alpha} < U \leq k^{-\alpha}) = 
k^{-\alpha}-(k+1)^{-\alpha}.
\end{equation}
For large $\alpha$, most of the probability mass is concentrated in the first few $k$, since $\PP(X>k) = (k+1)^{-\alpha}$. For large $k$, we can approximate $\phi_{\alpha}(k)$ by 
\begin{equation}
\label{eq:app}
\phi_{\alpha}(k) = k^{-\alpha}\Big[1-\frac{1}{(1+1/k)^{\alpha}}\Big] \simeq \frac{\alpha k^{-\alpha}}{k+\alpha}.
\end{equation}
This approximation is actually quite good already for small $k$ if $\alpha$ is not too large; when $\alpha=1$ it recovers the exact distribution. We see from (\ref{eq:app}) that the tail behavior of $\phi_{\alpha}(k)$ is analogous to that of the discrete Pareto (or Zipf) distribution \cite[Sec.~11.2.20]{univariate}
\begin{equation}
\phi_{\alpha}^{P}(k) = \frac{k^{-1-\alpha}}{\zeta(1+\alpha)}, \quad \alpha>0,\ k \geq 1,
\end{equation}
where $\zeta(z)$ is the usual zeta function. The continuous Student's $t$ distribution $t_{\nu}(x)$ displays the same tail behavior with the number of degrees of freedom $\nu$ playing the role of $\alpha$ \cite{lispre}. We refer to the real parameter $\alpha>0$ as the tail index of the distribution. 

The expectation of the random variable $X \sim \phi_{\alpha}$ is given by
\begin{equation}
\EE(X) = \sum_{k=1}^{\infty}k\phi_{\alpha}(k) = \zeta(\alpha) 
\end{equation}
and diverges if $\alpha \leq 1$, while its variance 
\begin{equation}
\label{eq:divar}
\var(X) = \EE(X^{2})-\EE(X)^{2} = 2\zeta(\alpha-1)-\zeta(\alpha)-\zeta(\alpha)^{2}
\end{equation}
diverges if $\alpha \leq 2$. These facts are relevant in our analyses, see Section~\ref{sec:facts}.

\begin{figure}[t]
\centering
\includegraphics[viewport= 0 10 550 330, scale=0.34, clip]{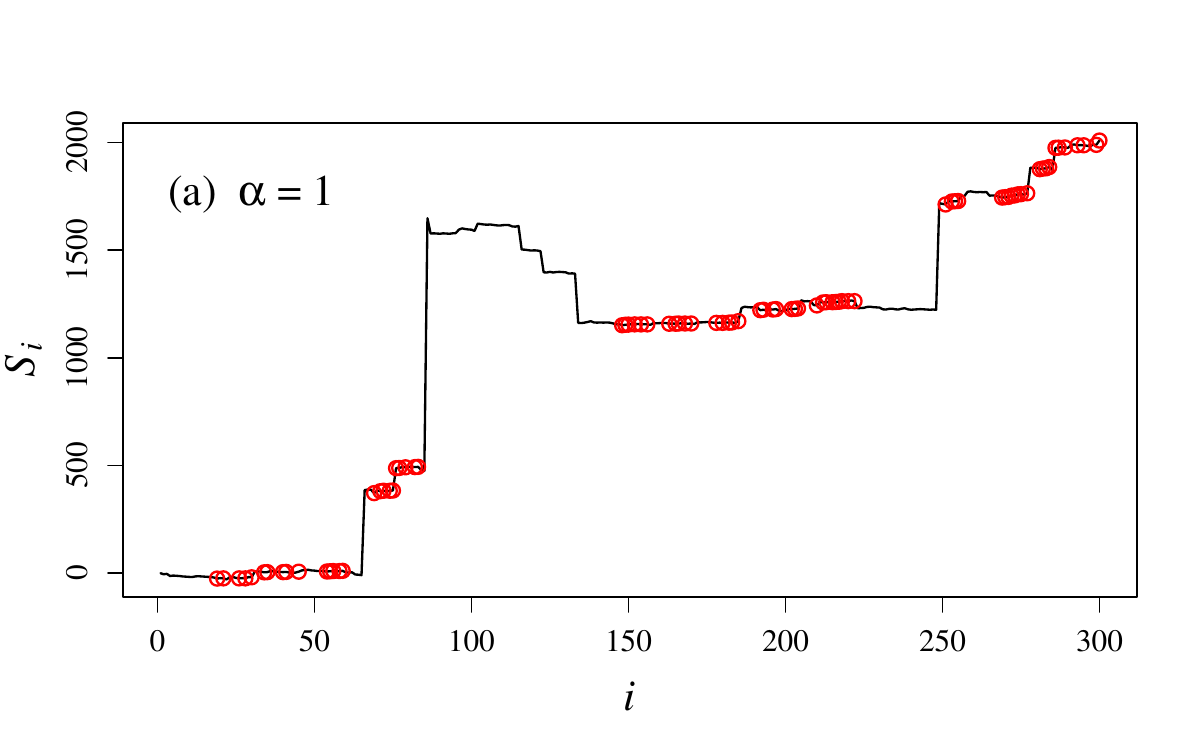} \hfill
\includegraphics[viewport= 0 10 550 330, scale=0.34, clip]{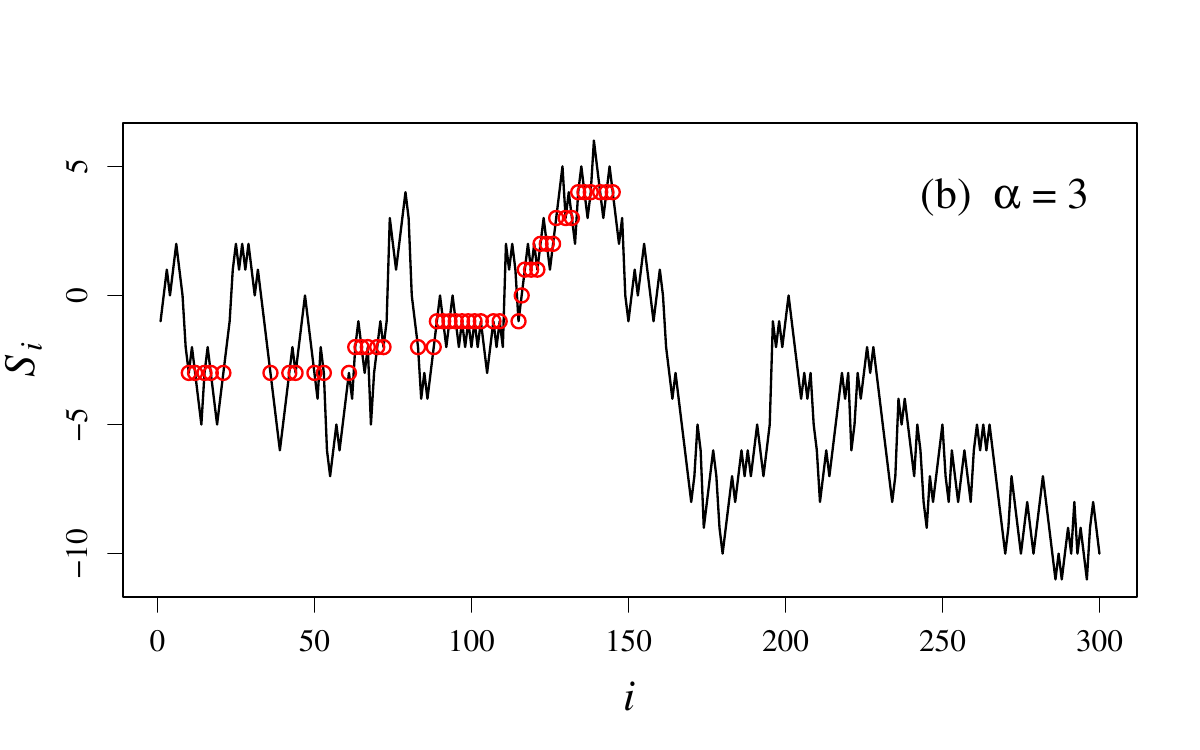} \\
\includegraphics[viewport= 0 10 550 350, scale=0.34, clip]{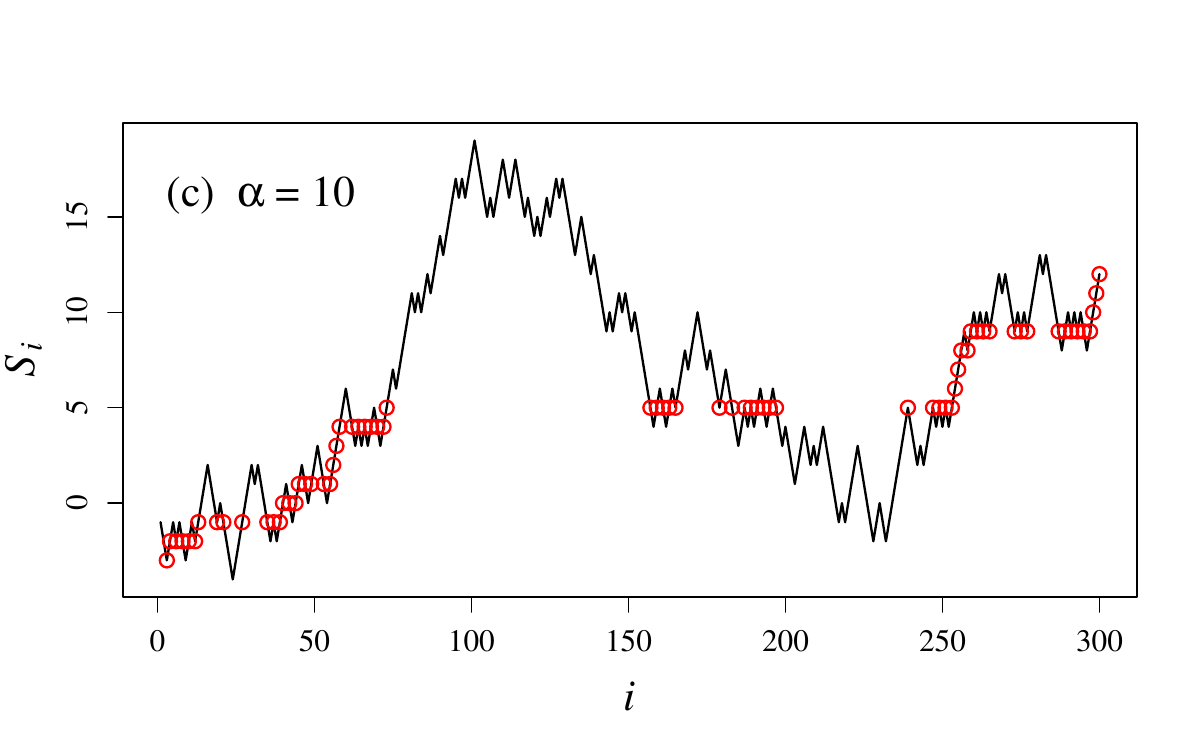} \hfill
\includegraphics[viewport= 0 10 550 350, scale=0.34, clip]{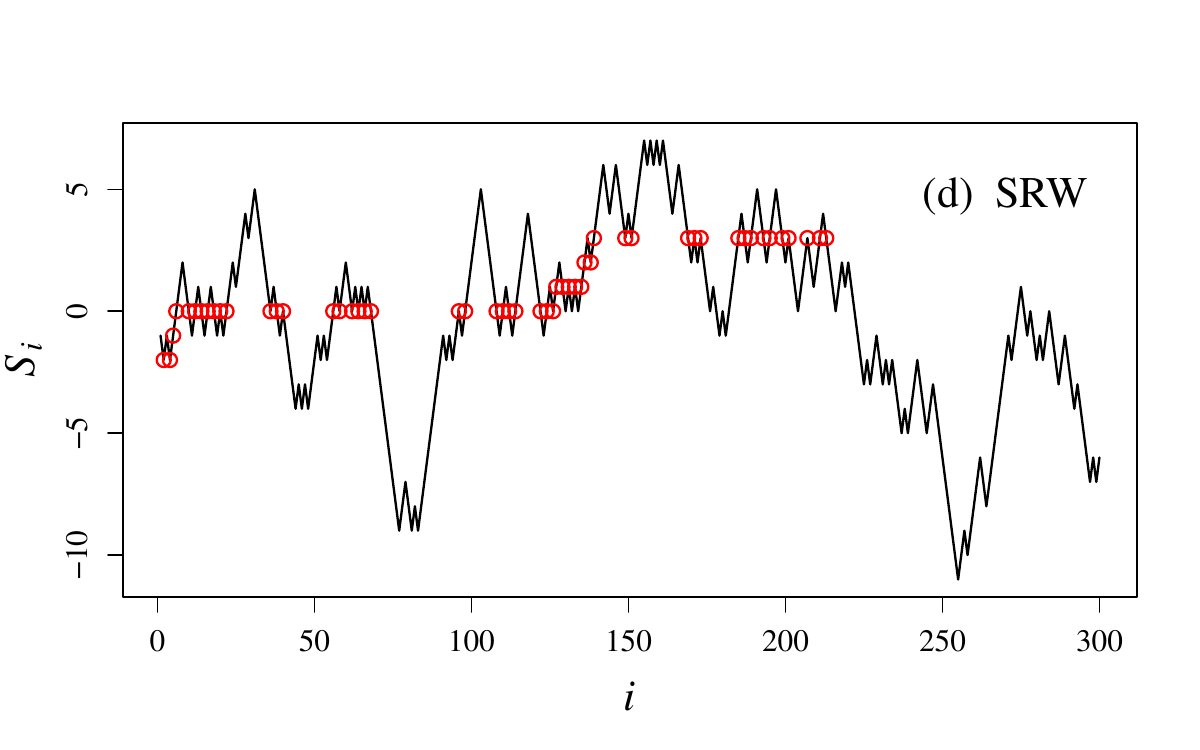}
\caption{Sample random walks (\ref{eq:sym}) and their associated weak LIS (circles) for different values of $\alpha$ (see Eq.~(\ref{eq:prb})) and a simple random walk (SRW). Note the different vertical scales. For $n=300$, the random walk with $\alpha=10$ and the simple random walk are virtually indistinguishable, since in this case the probability that some step increment or decrement of the walk is larger than $1$ becomes smaller than $10^{-3}$; in panel~(c), for instance, there is none.}
\label{fig:lis}
\end{figure}

The random variable $X$ defined above is strictly positive. To set up the random walk (\ref{eq:rw}), the increments $X_{i}$ must assume positive and negative values alike. We obtain such $X_{i}$ by multiplying each draw from $\phi_{\alpha}$ by a random sign $s_{i}$. The $n$-step discrete symmetric heavy tailed random walk on the integers then becomes
\begin{equation}
\label{eq:sym}
S_{0}=0,\ S_{i} = S_{i-1}+s_{i}X_{i},\ 1 \leq i \leq n,
\end{equation}
with the $X_{i}$ \iid\ $\phi_{\alpha}$ and the $s_{i} = \pm 1$ uniformly at random. The behavior of the variance of the signed $X_{i}$ is the same as in (\ref{eq:divar}), since $\var(sX)=\EE(X^{2})$ diverges when $\alpha \leq 2$ in the same way. In the infinite-variance range $0 < \alpha < 2$, the walks~(\ref{eq:sym}) may be viewed as lattice analogues of L\'{e}vy-flight-type random walks, although they are not $\alpha$-stable processes because the increments are integer-valued. Note that the larger the parameter $\alpha$, the more the heavy tailed random walk (\ref{eq:sym}) resembles a simple random walk with steps $\pm 1$. Figure~\ref{fig:lis} displays sample walks of length $n=300$ together with their weak LIS for $\alpha = 1$, $3$, $10$ and the simple random walk (SRW), illustrating how the typical jump amplitude shrinks and the trajectory becomes increasingly SRW-like as $\alpha$ grows. Quantitatively, in a finite random walk of length~$n$ with $X_{i} \sim \phi_{\alpha}$, whenever
\begin{equation}
\label{eq:srwlim}
\alpha > (1+o(1))\log_{2}{n}\ \Rightarrow\ \PP(X_{i}>1) < 1/n.
\end{equation}
Thus, for any finite walk length $n$, the heavy-tailed walk~(\ref{eq:sym}) reduces to the simple random walk with high probability when $\alpha$ is sufficiently large; the SRW can be regarded as the $\alpha \to \infty$ limiting case.

\section{\label{sec:facts}Known facts about the LIS of random walks}

In the mathematical literature, one can find the following rigorous results on the LIS of symmetric random walks. When the distribution of increments $\phi_{\alpha}$ is a symmetric \textit{continuous} distribution of finite variance, then for all $\eps>0$ and sufficiently large~$n$ the expected length of the LIS is bounded as \cite{angel}
\begin{equation}
\label{eq:finite}
c\sqrt{n} \leq \EE(L_{n}) \leq n^{1/2+\eps}
\end{equation}
for some $c>0$. Note that the lower bound $\EE(L_{n}) \geq \sqrt{n}$ (in fact, $L_{n} \geq \sqrt{n}$) follows from the Erd\H{o}s-Szekeres theorem \cite{hammersley}. Note also that for continuous distributions of increments, the distinction between the weak ($S_{i_{1}} \leq \cdots \leq S_{i_{L}}$) and the strict ($S_{i_{1}} < \cdots < S_{i_{L}}$) LIS is immaterial. For the simple, \textit{discrete} random walk with step increments $\pm 1$, the bounds for the expected length of the \textit{weak} LIS become
\begin{equation}
\label{weaklis}
c\sqrt{n}\mku\log{n} \leq \EE(\text{weak-}L_{n}) \leq n^{1/2+\eps},
\end{equation}
leaving only a slowly varying factor undetermined in this case. In \cite{angel}, the authors further note that the arguments leading to the additional logarithmic factor in the lower bound in (\ref{weaklis}) seem to hold also for random walks with step increments $\pm 1$, \ldots, $\pm k$ for finite, not very large $k$. When the symmetric continuous $\phi_{\alpha}$ has infinite variance, the bounds on $\EE(L_{n})$ become \cite{pemantle}
\begin{equation}
\label{eq:infinite}
n^{\beta_{0}-\eps} \leq \mathbb{E}(L_{n}) \leq n^{\beta_{1}+\eps},
\end{equation}
where $\beta_{0} \simeq 0.690$ is the positive solution of $x+2^{-1-x}=1$ and $\beta_{1} \simeq 0.815$ is obtained from the numerical solution of an implicit equation involving a non-elementary integral; these exponents are not sharp. Presumably this upper bound (which relies on general arguments about the walk's range) and lower bound (which uses a specific construction) also apply to the discrete case.

\begin{figure}[t]
\centering
\includegraphics[viewport=0 10 480 460, scale=0.34, clip]{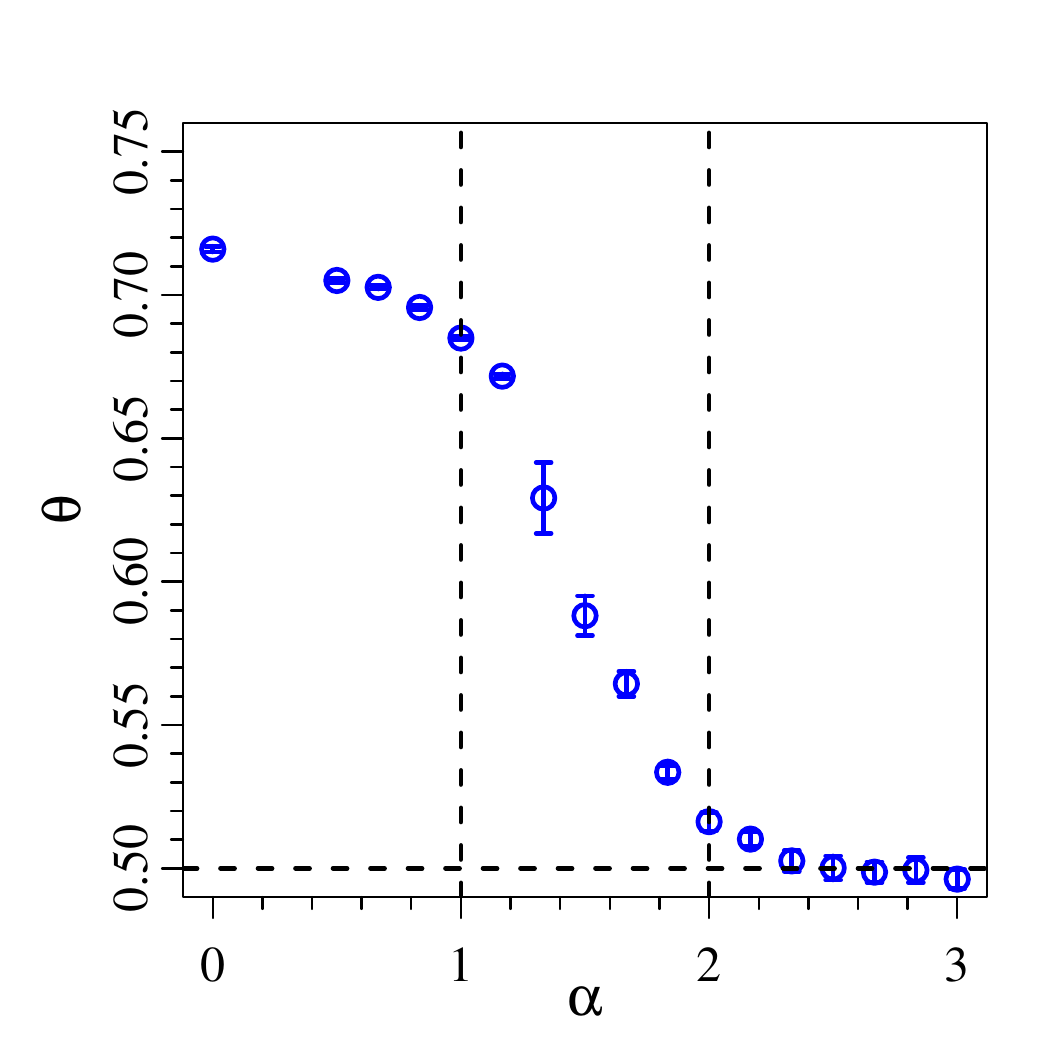}
\caption{Leading exponent $\theta$ (see (\ref{eq:heavy})) in the asymptotic behavior of the LIS of random walks with symmetric continuous heavy tailed distribution of increments $\phi_{\alpha}(\abs{x} \gg 1) \sim \abs{x}^{-1-\alpha}$ in the range $0 < \alpha \leq 3$. The point at $\alpha=0$ represents the value of $\theta$ found for the LIS of the ultra-fat random walk. When $\alpha \geq 2$ the values of $\theta$ approach $0.5$ and the asymptotic behavior of the LIS acquires a logarithmic factor, see (\ref{eq:short}). The data plotted are a compilation of results from previous numerical studies \cite{lisjpa,hartmann,lispre,ultrafat}; the present work extends this picture from the continuous-walk to the discrete-walk setting (see Tables~\ref{tab:adjI}--\ref{tab:adjIIb} below).}
\label{fig:theta}
\end{figure}

On the non-rigorous side, numerical simulations showed that when $\phi_{\alpha}$ is continuous of finite variance---for instance, normal or Student's $t_{\nu}$ of $\nu>2$ degrees of freedom---, the sample average $\langle{L_{n}}\rangle$ behaves for large~$n$ as
\begin{equation}
\label{eq:short}
\langle{L_{n}}\rangle \sim \sqrt{n}\mku(a+b\log{n})
\end{equation}
with $a \simeq b \simeq 0.36$ \cite{lisjpa,hartmann,lispre}, while for a continuous heavy tailed $\phi_{\alpha}$, $\langle{L_{n}}\rangle$ behaves for large~$n$ like
\begin{equation}
\label{eq:heavy}
\langle{L_{n}}\rangle \sim n^{\theta}
\end{equation}
with an exponent $0.5 < \theta \lesssim 0.716$ depending on the heaviness of $\phi_{\alpha}$---the heavier the $\phi_{\alpha}$, the larger the $\theta$. The largest $\theta = 0.716 \pm 0.002$ was found for the LIS of ultra-fat tailed random walks, which have the heaviest possible distribution of increments, without any finite positive moments. Figure~\ref{fig:theta} summarizes the behavior of the leading exponent $\theta$ in the asymptotic behavior of the LIS of random walks with continuous symmetric heavy tailed distribution of increments $\phi_{\alpha}(\abs{x} \gg 1) \sim \abs{x}^{-1-\alpha}$ \cite{lisjpa,hartmann,lispre,ultrafat}. The value at $\alpha=0$ in Figure~\ref{fig:theta} stands for the exponent found for the LIS of symmetric ultra-fat tailed (also discrete) random walks; see \cite{pemantle,ultrafat} for a discussion.

The $\log{n}$ term in (\ref{eq:short}), with $\phi_{\alpha}$ continuous, came somewhat as a surprise. While the upper bound $n^{1/2+\eps}$ in (\ref{eq:finite}) leaves room for a $(\log{n})^p$ term with some $p>0$, the lower bound there is sharp. Rigorously, such logarithmic behavior has been established only for the weak LIS of simple random walks, with a discrete distribution of increments; see (\ref{weaklis}). In such random walks, the repeated visitation of lattice levels creates plateaus that can be exploited by weakly increasing subsequences, since equal values are allowed in the subsequence. The presence of such plateaus is therefore expected to modify the combinatorics of the longest increasing subsequences and may account for the logarithmic corrections observed in the discrete case. Our study with discrete distributions of increments provides direct evidence on this point. This mechanism has no counterpart in random walks with continuous increments, where ties occur with probability zero. It remains unclear whether the appearance of this logarithmic term reflects a genuine property of random walks with continuous increments or is an artifact of the intrinsic discreteness of the numerical simulations.


\section{\label{sec:lis}Exploratory analysis of the LIS}

For each of $13$ walk lengths in the range $10^{4} \leq n \leq 10^{8}$ and distribution of increments with tail indices $1/2 \leq \alpha \leq 10$, we generate $10{,}000$ random walks (\ref{eq:sym}) and compute the length $L_{n}$ of their weak LIS. We have also simulated the simple random walk to check whether and how the weak LIS of heavy tailed random walks converge to the weak LIS of the simple random walk.

\begin{figure}[t]
\centering
\includegraphics[viewport= 10 10 480 340, scale=0.40, clip]{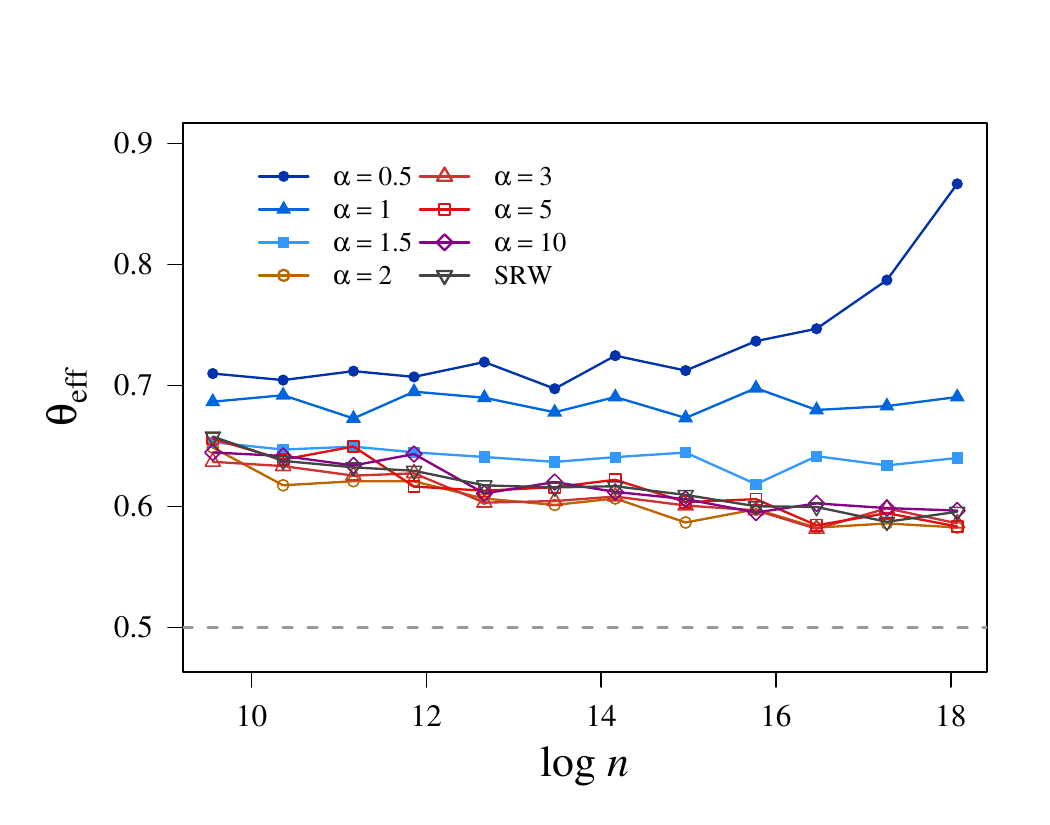}
\caption{Effective exponent $\theta_{\text{eff}}(n) = \Delta\log\langle L_{n}\rangle /\Delta\log n$ of the weak LIS of discrete heavy tailed random walks (\ref{eq:sym}) as a function of $\log n$, evaluated at the midpoints of consecutive pairs of walk lengths. For $\alpha = 1$, $\theta_{\text{eff}}$ is approximately constant, consistent with a pure power law $\langle L_{n}\rangle \sim n^{\theta}$; for $\alpha = 1/2$, $\theta_{\text{eff}}$ is roughly constant up to $n \approx 10^{7}$ but drifts upward at the largest walk lengths (see Section~\ref{susub:caution}). For $\alpha \geq 2$ and the simple random walk (SRW), $\theta_{\text{eff}}$ decreases systematically, indicating the presence of a subleading logarithmic correction. The dashed line marks $\theta = 1/2$.}
\label{fig:effexp}
\end{figure}

We are interested in extracting the leading asymptotic behavior of the sample average (over the $10{,}000$ Monte Carlo realizations) $\langle{L_{n}}\rangle$ from the data as a function of the tail index $\alpha$. If $\langle{L_{n}}\rangle$ followed a pure power law $\langle{L_{n}}\rangle \sim an^{\theta}$, then $\log\langle{L_{n}}\rangle$ would be exactly linear in $\log n$ with slope $\theta$. Figure~\ref{fig:effexp} shows the effective (local) exponent
\begin{equation}
\label{eq:effexp}
\theta_{\text{eff}}(n) = \frac{\Delta\log{\langle{L_{n}}\rangle}}{\Delta\log{n}},
\end{equation}
computed from the data as a function of $\log n$ for each $\alpha$. For $\alpha = 1$, the effective exponent is approximately constant across the whole range of~$n$, consistent with a pure power law. For $\alpha = 1/2$, $\theta_{\text{eff}}(n)$ is also roughly constant up to $n \approx 10^{7}$ but then drifts upward at the largest walk lengths, reflecting the extreme volatility of this tail index (in which not even the first moment of the increments is defined); we return to this point in detail in Section~\ref{susub:caution}. For $\alpha \geq 2$ and the simple random walk, however, $\theta_{\text{eff}}(n)$ decreases systematically with~$n$, revealing a downward trend in the $\log\langle{L_{n}}\rangle$ vs.\ $\log n$ plot that is incompatible with a simple power law. This behavior is consistent with the ansatz $\langle{L_{n}}\rangle \sim \sqrt{n}\mku(a+b\log n)$, for which (replacing discrete differences by continuous derivatives in (\ref{eq:effexp}))
\begin{equation}
\label{eq:tends}
\theta_{\text{eff}}(n) = \frac{1}{2}+\frac{b}{a+b\log n} \to \frac{1}{2}
\end{equation}
from above as $n \to \infty$. The case $\alpha = 3/2$ displays an intermediate behavior.

Based on the existing theoretical ground and this initial analysis, we propose two different models to describe the data, which we call Model~I and Model~II, and later select which better does the job. In Model~I, we fit the model 
\begin{equation}
\label{eq:adjI}
f(n;\theta,a,b) = n^{\theta}(a+b\log{n})
\end{equation}
to the data irrespective of the heavy tail index $\alpha$. In Model~II, we fit models according to the ansätze (\ref{eq:short})--(\ref{eq:heavy}),
\begin{equation}
\label{eq:adjII}
f(n;\theta,a,b) = 
\begin{cases}
an^{\theta}, & \text{for the data with } \alpha < 2; \\
\sqrt{n}\mku(a+b\log{n}), & \text{for the data with } \alpha \geq 2.
\end{cases}
\end{equation}
In all cases we look for the parameters $\theta$, $a$, and $b$ that minimize the residual sum of squares
\begin{equation}
\label{eq:lsquares}
\RSS(\theta,a,b) = \sum_{n}\big[\langle{L_{n}}\rangle-f(n;\theta,a,b) \big]^{2}.
\end{equation} 
Both models are consistent with $a=1$ at $n=1$, though the fits are driven by data at much larger~$n$ and should not be expected to extrapolate precisely to $n=1$. As we shall see, in Model~I we indeed find $a \approx 1$ in most cases, but in Model~II we obtain a more diverse set of values (not much off $1$, though). This is, of course, due to the noisy and finite nature of the data. 

\subsection{\label{sec:adjI}Model~I}

In Model~I, we evaluate the fits by direct numerical evaluation of $\RSS(\theta,a,b)$ on a grid of discretized points $(\theta,a,b)$. Based on previous studies \cite{lisjpa,lispre}, we scan the parameters in the ranges $0.5 \leq \theta \leq 0.8$, $0 \leq a \leq 2$, and $0 \leq b \leq 1$ in steps of size $0.01$, generating $31 \times 201 \times 101 = 629,\!331$ evaluation points for $\RSS(\theta,a,b)$. The range of $\theta$ is motivated by the rigorous bounds (\ref{eq:infinite}), which constrain the exponent to lie below $\beta_{1} \approx 0.815$, and by the empirical upper bound $\theta_{\text{ultra}} \approx 0.716$ found for the ultra-fat tailed random walk \cite{ultrafat}. The largest best-fit exponent, $\bar{\theta} \approx 0.749$ found for $\alpha=1/2$ (see Table~\ref{tab:adjI}), lies comfortably within the grid, and the remaining parameters are similarly far from their respective boundaries.

To estimate the parameters, we select all grid points $(\theta_{i},a_{i},b_{i})$ whose $\RSS_{i} \leq \sqrt{10}\mku\RSS_{\text{min}}$, \ie, those producing residual sums of squares within the same order of magnitude of the global minimum. This typically yields a few dozen or fewer fits to consider. The selected estimates are then averaged with weights proportional to $1/\sqrt{\RSS}$. For example, for the exponent $\theta$ we obtain the weighted estimate $\overline{\theta} = \sum_{i}w_{i}\theta_{i} \pm s_{\overline{\theta}}$ with
\begin{equation}
\label{eq:weights}
w_{i} = \frac{1/\sqrt{\RSS_{i}(\theta_{i},a_{i},b_{i})}}{\sum\limits_{j}1/\sqrt{\RSS_{j}(\theta_{j},a_{j},b_{j})}} \quad \text{and} \quad s_{\overline{\theta}} = s_{\theta}\sqrt{\textstyle\sum\limits_{i}w_{i}^{2}}\,,
\end{equation}
where $i$ and $j$ run over the selected fits and $s_{\theta}$ is the (corrected) sample standard deviation of $\theta$ calculated from the $\theta_{i}$.
   
A few remarks about this procedure are in order. First, the grid spacing of $0.01$ sets an intrinsic resolution limit on the parameter estimates; some of the uncertainties quoted in Table~\ref{tab:adjI} are comparable to this limit. Second, the uncertainty $s_{\overline{\theta}}$ should be regarded as an estimate of the precision of the weighted average rather than a rigorous confidence interval, since the selected grid points are spatially correlated and the supposition that they are \iid\ is only approximately valid. Third, the residual sum of squares (\ref{eq:lsquares}) treats all walk lengths equally. Since $\langle{L_{n}}\rangle$ spans several orders of magnitude across the range of~$n$ considered, the residuals at the largest walk lengths tend to dominate the RSS, effectively giving more weight to the asymptotic regime. While this is arguably desirable when extracting the asymptotic behavior, it does mean that the fits are less sensitive to the data at smaller~$n$. The estimated parameters thus obtained appear in Table~\ref{tab:adjI}. We do not quote (or directly employ) the coefficients of determination $R^{2}$ of the fits because they are all very high, $R^{2}>0.99$ or higher in all cases.

\begin{table}[h]
\caption{Best weight-averaged estimated parameters of a least-squares fit to the data according to Model~I (\ref{eq:adjI}), irrespective of the heavy tail index $\alpha$. The last column indicates the number of fits contributing to the averages in each case according to our criterion for inclusion. Uncertainties in the last digits of the estimates are indicated within parentheses.}
\centering
\begin{tabular}{c|cccc}
\hline
$\alpha$ & $\overline{\theta}$ & $\overline{a}$ & $\overline{b}$ & \#\,fits \\ 
\hline
$1/2$      & $0.749(3)$  &  $0.154(17)$ &  $0.025(2)$  & $56$ \\
$1$        & $0.658(9)$  &   $0.72(13)$ &  $0.06(2)$   & $8$  \\
$3/2$      & $0.605(5)$  &   $1.13(14)$ &  $0.10(2)$   & $13$ \\[2pt] 
\hline \\[-9pt]
$2$        & $0.533(2)$  &   $0.7(2)$   &  $0.40(3)$   & $6$  \\ 
$3$        & $0.543(1)$  &   $1.12(10)$ &  $0.36(1)$   & $36$ \\
$5$        & $0.539(1)$  &   $0.97(7)$  &  $0.48(1)$   & $76$ \\
$10$       & $0.547(2)$  &   $0.9(2)$   &  $0.44(3)$   & $9$  \\ 
SRW        & $0.543(1)$  &   $0.93(8)$  &  $0.47(1)$   & $49$ \\ 
\hline
\end{tabular}
\label{tab:adjI}
\end{table}

We see from Table~\ref{tab:adjI} that, as expected, the leading exponent $\theta$ in the asymptotic behavior of $\langle{L_{n}}\rangle$ changes around $\alpha=2$: while it increases when $\alpha \searrow 0$, it stagnates at $\theta \simeq 0.54$ for $\alpha \geq 2$, in accordance with previous observations for random walks with step increments of finite variance \cite{lisjpa,hartmann,lispre,ultrafat}. We also see from Table~\ref{tab:adjI} that for $\alpha<2$, parameter $a$ is much larger than parameter $b$, indicating a relatively minor importance of the subleading term proportional to $\log{n}$ in the model, somewhat validating the ansatz (\ref{eq:heavy}) in this case.

In \cite{ultrafat}, we found $\theta_{\text{ultra}} = 0.716 \pm 0.002$ for the symmetric ultra-fat tailed random walk, a discrete random walk on the integers which can be thought of as equivalent to a symmetric $\alpha$-stable random walk with $\alpha \approx 0$, although we do not know precisely to which $\alpha$-stable random walk the finite-size ultra-fat tailed random walk corresponds; see Figure~\ref{fig:theta} and \cite{pemantle,ultrafat} for a discussion. We would then expect that the exponent $\theta$ found for the discrete heavy tailed random walk on the integers with tail index $\alpha=1/2$ would be smaller than $\theta_{\text{ultra}}$, not greater. It must be noticed, however, that the longest random walk studied in \cite{ultrafat} had $n=2^{18}$ steps, almost $400$ times smaller than the longest random walks studied here. It is possible that the analysis of the leading exponent $\theta$ of larger ultra-fat tailed random walks would get closer to saturating the bound $\theta \leq 0.815$ in (\ref{eq:infinite}). The numerical calculation of the LIS of ultra-fat tailed random walks as defined in \cite{pemantle,ultrafat} is quite a CPU-time demanding task, mainly because it involves the linear search of a certain maximum index in the random walk when computing its LIS (cf.\ \cite[Eq.~(4)~{\&\ seq.}]{ultrafat} for details), which greatly hampers the calculation of thousands of LIS of very long random walks, as required to obtain good statistics.

\subsection{\label{sec:adjII}Model~II} 
 
In Model~II, the ansätze (\ref{eq:adjII}) can be reduced to simple linear regressions. When $\alpha < 2$, we fit $\log\langle{L_{n}}\rangle = \log{a}+\theta\log{n}$, a linear model in $\log{n}$ with slope $\theta$ and intercept $\log{a}$. When $\alpha \geq 2$, we fit $\langle{L_{n}}\rangle/\sqrt{n} = a+b\log{n}$, a linear model in $\log{n}$ with slope $b$ and intercept $a$. The linear regressions were performed by means of the R (version 4.5.3) linear model utility \texttt{lm(Y\,$\sim$\,X)}, with $Y$ given either by $\log\langle{L_{n}}\rangle$ or $\langle{L_{n}}\rangle/\sqrt{n}$ and $X=\log{n}$.

Note that when we do a linear regression for $\log\langle{L_{n}}\rangle$, we are implicitly assuming that the errors are multiplicative or, equivalently, that $\log\langle{L_{n}}\rangle$ has approximately constant variance across~$n$. Otherwise, the linear fit for $\langle{L_{n}}\rangle/\sqrt{n}$ assumes additive errors. The error assumptions implicit in these linear regressions are addressed more carefully in Section~\ref{sec:anova}. 
 
\subsubsection{Tail index $\alpha < 2$: pure power law}

The estimated parameters for $\alpha < 2$ are shown in Table~\ref{tab:adjIIa}. The exponent $\theta$ decreases with increasing $\alpha$, from $\theta \approx 0.73$ at $\alpha=1/2$ to $\theta \approx 0.64$ at $\alpha=3/2$, in qualitative agreement with Model~I (see Table~\ref{tab:adjI}). The values of $\theta$ differ quantitatively from those of Model~I, however, because Model~I includes the subleading term $b\log{n}$ which absorbs part of the scaling and shifts $\theta$. The effective exponent plot in Figure~\ref{fig:effexp} supports the pure power law for $\alpha \leq 1$, where $\theta_{\text{eff}}(n)$ is approximately constant. At $\alpha = 3/2$, a mild downward drift in $\theta_{\text{eff}}(n)$ is visible, suggesting that a logarithmic correction may already be relevant at this value of $\alpha$. We note that for the ultra-fat tailed random walk, the best fits were also obtained with $b=0$ \cite{ultrafat}.
 
\begin{table}[h] 
\caption{Estimated parameters of a least-squares fit of Model~II to the data with $\alpha < 2$, according to (\ref{eq:adjII}). Uncertainties in the last digits of the estimates are indicated within parentheses.} 
\centering 
\begin{tabular}{c|cc}
\hline 
$\alpha$ & $\theta$ & $a$ \\ \hline 
$1/2$ & $0.726(5)$ & $0.76(5)$ \\
$1$& $0.685(0)$ & $1.059(6)$ \\
$3/2$ & $0.640(1)$ & $1.54(2)$ \\ \hline 
\end{tabular} 
\label{tab:adjIIa}
\end{table}

\subsubsection{Tail index $\alpha \geq 2$: logarithmic correction to $\sqrt{n}$}

The estimated parameters for $\alpha \geq 2$ and the simple random walk are shown in Table~\ref{tab:adjIIb}. The intercept $a$ is negative in all cases. This does not imply a negative LIS; $a$ is merely an adjustment constant in the linear model, and the fitted values $\langle{L_{n}}\rangle/\sqrt{n} = a+b\log{n}$ remain positive for all~$n$ in the data range. The slope $b$ increases monotonically from $b \approx 0.98$ at $\alpha=2$ to $b \approx 1.50$ at $\alpha=10$, at which point it matches the simple random walk, confirming that the discrete heavy tailed random walk converges to the simple random walk for large $\alpha$. 
 
\begin{table}[h]
\caption{Estimated parameters of a least-squares fit of Model~II to the data with $\alpha \geq 2$ and the simple random walk (SRW), according to (\ref{eq:adjII}). The exponent $\theta=1/2$ is fixed by the model. Uncertainties in the last digits of the estimates are indicated within parentheses.}
\centering 
\begin{tabular}{c|cc}
\hline
$\alpha$ & $a$ & $b$ \\ \hline
$2$& $-3.6(2)$ & $0.98(2)$ \\
$3$& $-4.8(3)$ & $1.16(2)$ \\ 
$5$& $-6.6(4)$ & $1.43(3)$ \\
$10$ & $-7.2(5)$ & $1.50(3)$ \\ 
SRW& $-7.1(5)$ & $1.50(3)$ \\ \hline
\end{tabular} 
\label{tab:adjIIb}
\end{table} 

The coefficients $a$ and $b$ in Table~\ref{tab:adjIIb} are significantly different from those found for random walks with continuous, finite variance distributions of increments, for which $a \approx b \approx 0.36$ \cite{lisjpa,hartmann,lispre}. This discrepancy is not unexpected. For continuous distributions, the logarithmic term came as a surprise, since the rigorous lower bound $\EE(L_{n}) \geq c\sqrt{n}$ in (\ref{eq:finite}) does not require it. For discrete distributions, on the other hand, the stronger lower bound $\EE(\text{weak-}L_{n}) \geq c\sqrt{n}\mku\log{n}$ in (\ref{weaklis}) establishes the logarithmic correction on rigorous grounds \cite{angel}. The much larger values of $b$ (and correspondingly more negative $a$) found here reflect the enhanced role of the $\log{n}$ term in the discrete setting, where the weak LIS benefits from the flat segments that are inherent to integer-valued random walks.

\subsection{\label{sec:compare}A comparison of Models I and II}

It is natural to ask which model, I or II, better describes the data across the different regimes of $\alpha$. To address this, we employ two complementary diagnostic tools that do not rely on assumptions about the error structure: ratio plots and a stability analysis of the fitted exponent $\theta$.

In the ratio plots, we divide $\langle{L_{n}}\rangle$ by the leading term predicted by each model and look for systematic trends in the remainder. If the model is correct, the ratio should be approximately constant. In the stability analysis, we fit a pure power law $\langle{L_{n}}\rangle = an^{\theta}$ to the data for each $\alpha$, progressively dropping the smallest walk lengths and retaining only the data with $n \geq n_{\min}$. If $\theta$ is stable as $n_{\min}$ increases, the power law is a good description of the asymptotic behavior. If $\theta$ drifts systematically, a subleading correction is being absorbed into the exponent.

\subsubsection{Tail index $\alpha \geq 2$ and the simple random walk}

\begin{figure}[t]
\centering
\includegraphics[viewport= 5 10 480 460, scale=0.36, clip]{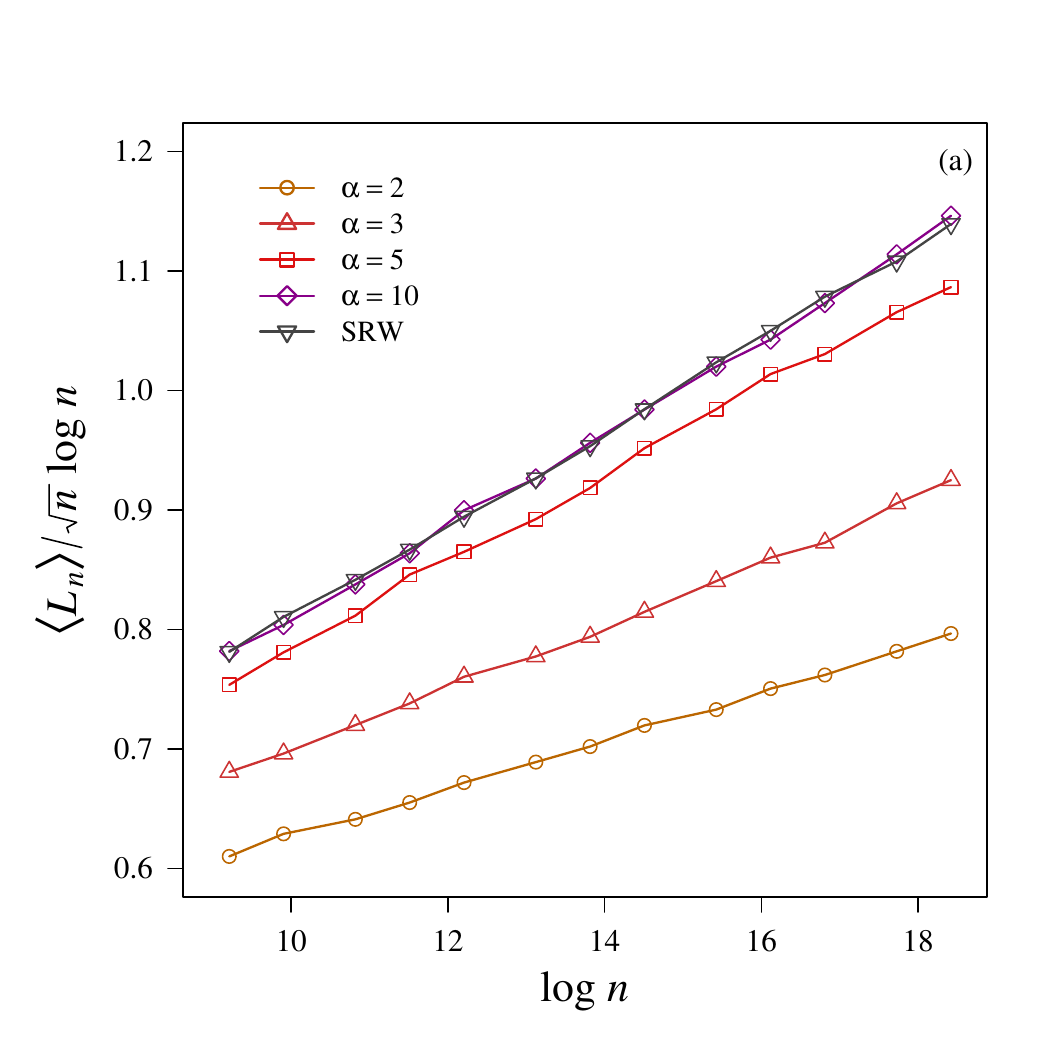} \hspace{1em} 
\includegraphics[viewport= 5 10 480 460, scale=0.36, clip]{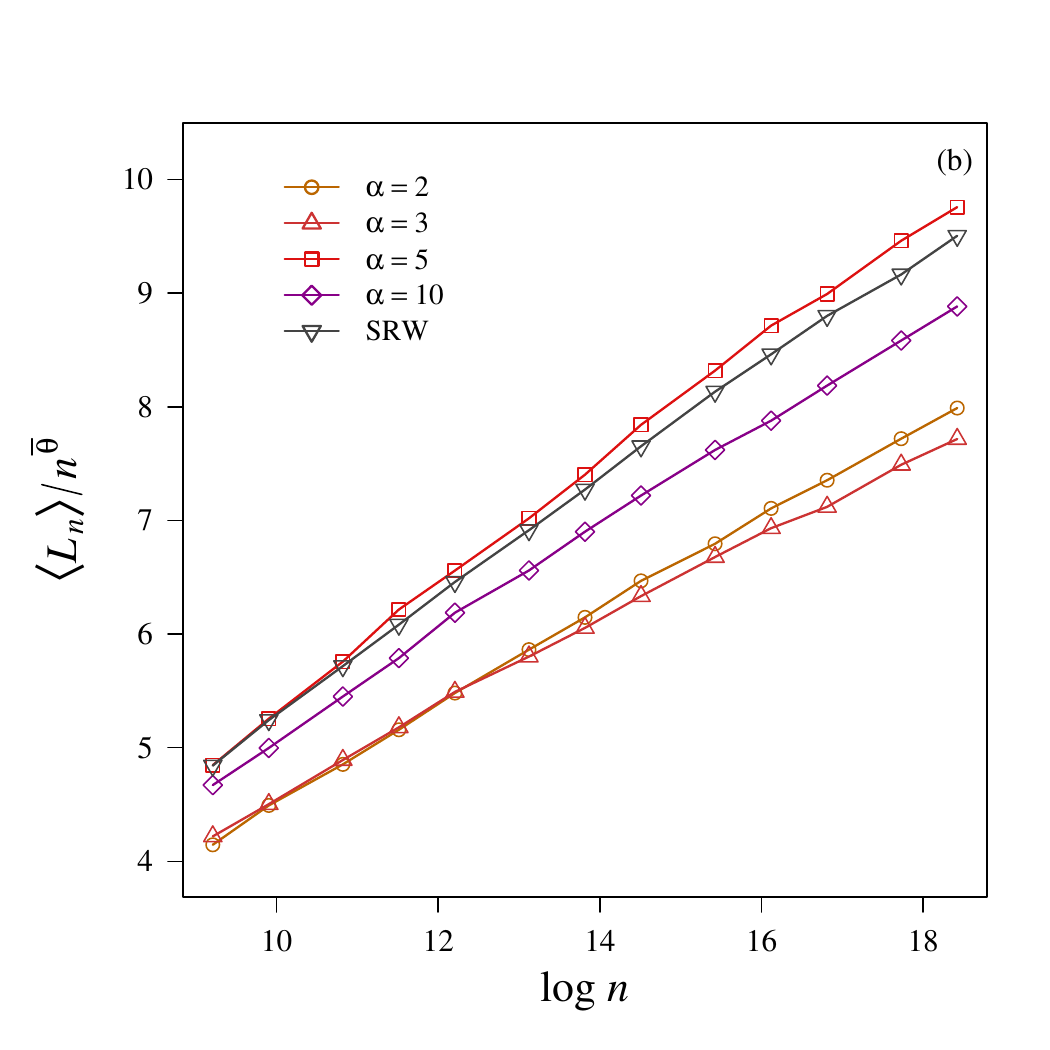}
\caption{Ratio plots for $\alpha \geq 2$ and the simple random walk (SRW). (a)~Ratio $\langle{L_{n}}\rangle/(\sqrt{n}\mku\log{n})$ as a function of $\log{n}$. The monotonic increase indicates that the asymptotic regime $\langle{L_{n}}\rangle \sim b\sqrt{n}\mku\log{n}$ has not yet been reached; the intercept $a$ in Model~II (Table~\ref{tab:adjIIb}) is still contributing at these values of~$n$. (b)~Ratio $\langle{L_{n}}\rangle/n^{\overline{\theta}}$ as a function of $\log{n}$, where $\overline{\theta}$ is the Model~I exponent from Table~\ref{tab:adjI} for each $\alpha$. The approximately linear growth with $\log{n}$ reveals a residual logarithmic correction not captured by the power law alone.}
\label{fig:ratio_finite}
\end{figure}

Figure~\ref{fig:ratio_finite}(a) shows the ratio $\langle{L_{n}}\rangle/(\sqrt{n}\mku\log{n})$ as a function of $\log{n}$ for all $\alpha \geq 2$ and the simple random walk. If the data obeyed a simple $\sqrt{n}\mku\log{n}$ scaling, these ratios would converge to a constant. Instead, they increase systematically with $\log{n}$. This growth is consistent with Model~II, in which $\langle{L_{n}}\rangle/\sqrt{n} = a+b\log{n}$ and the ratio becomes $a/\log{n}+b$, which converges to $b$ from below---a very slow approach. The data have not yet reached the asymptotic plateau, but the trend is clear.

Figure~\ref{fig:ratio_finite}(b) shows the ratio $\langle{L_{n}}\rangle/n^{\overline{\theta}}$ using Model~I's fitted $\overline{\theta} \simeq 0.54$ for the same values of $\alpha$. The ratios grow approximately linearly in $\log{n}$, revealing a residual logarithmic trend that the power law alone cannot absorb. This confirms the need for the subleading $\log{n}$ term in the model.

\begin{figure}[h]
\centering
\includegraphics[viewport= 10 10 480 460, scale=0.36, clip]{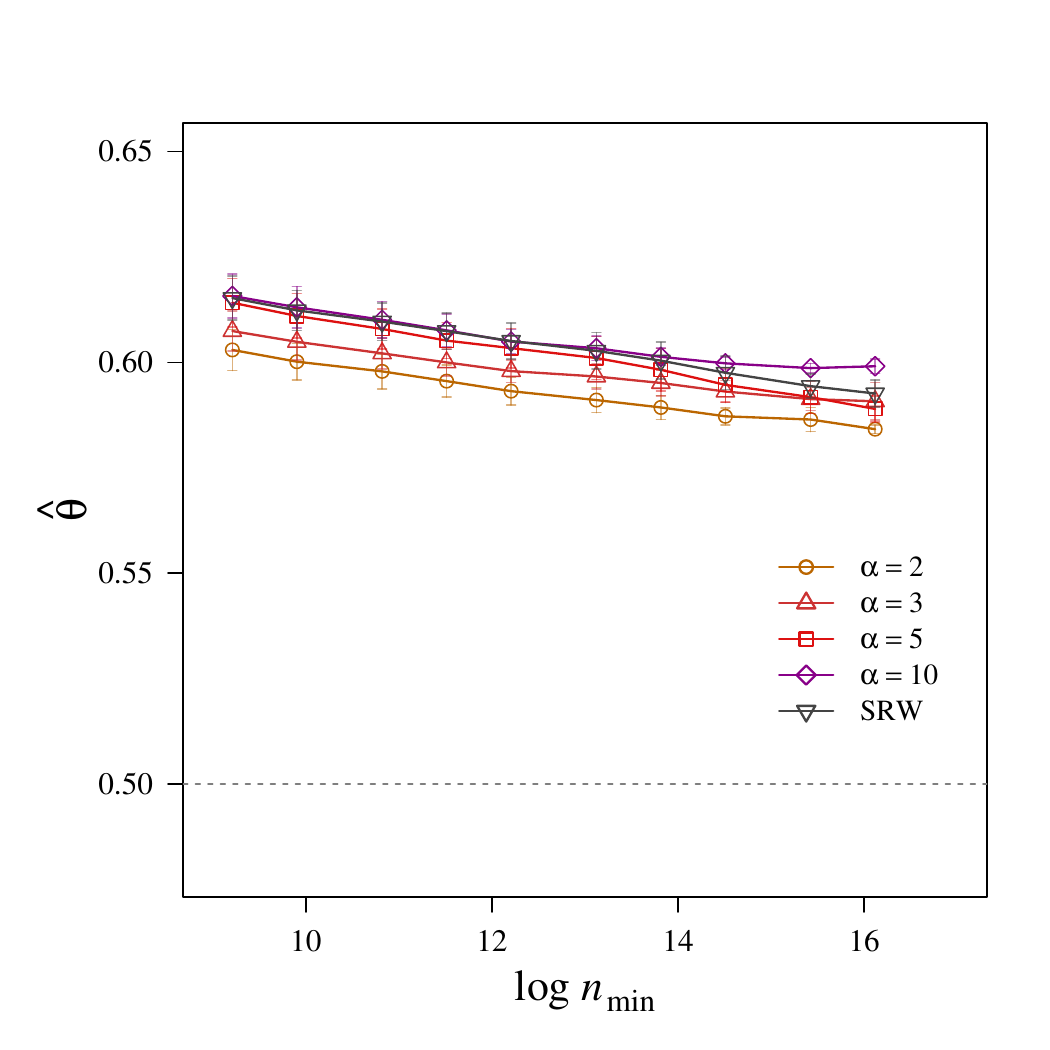}
\caption{Stability analysis for $\alpha \geq 2$ and the SRW. The estimated exponent $\hat{\theta}$ of a pure power law fit to the data with $n \geq n_{\min}$ is plotted as a function of $\log{n_{\min}}$. The systematic downward drift toward $\theta = 1/2$ (dashed line) indicates that the elevated $\hat{\theta}$ obtained from fitting all data points is an artifact of the logarithmic correction. Error bars indicate $\pm 2$ standard errors.}
\label{fig:stab_finite}
\end{figure}

The stability analysis, shown in Figure~\ref{fig:stab_finite}, corroborates this picture. When a pure power law is fitted to successively restricted data with $n \geq n_{\min}$, the estimated $\hat{\theta}$ decreases monotonically for all $\alpha \geq 2$ and the simple random walk, drifting from $\hat{\theta} \approx 0.60$--$0.62$ down to $\hat{\theta} \approx 0.58$--$0.60$ as the smallest walk lengths are removed. The drift is systematic, consistent across all values of $\alpha$, and clearly directed toward $\theta=1/2$. That $\hat{\theta}$ has not yet reached $1/2$ is expected: the convergence of the effective exponent is logarithmically slow when the true form involves a $\log{n}$ correction. These findings support Model~II for $\alpha \geq 2$, in which $\theta = 1/2$ exactly and the growth above $\sqrt{n}$ is entirely due to the logarithmic term.

\subsubsection{Tail index $\alpha < 2$}
\label{susub:caution}

\begin{figure}[t]
\centering
\includegraphics[viewport= 10 10 480 460, scale=0.36, clip]{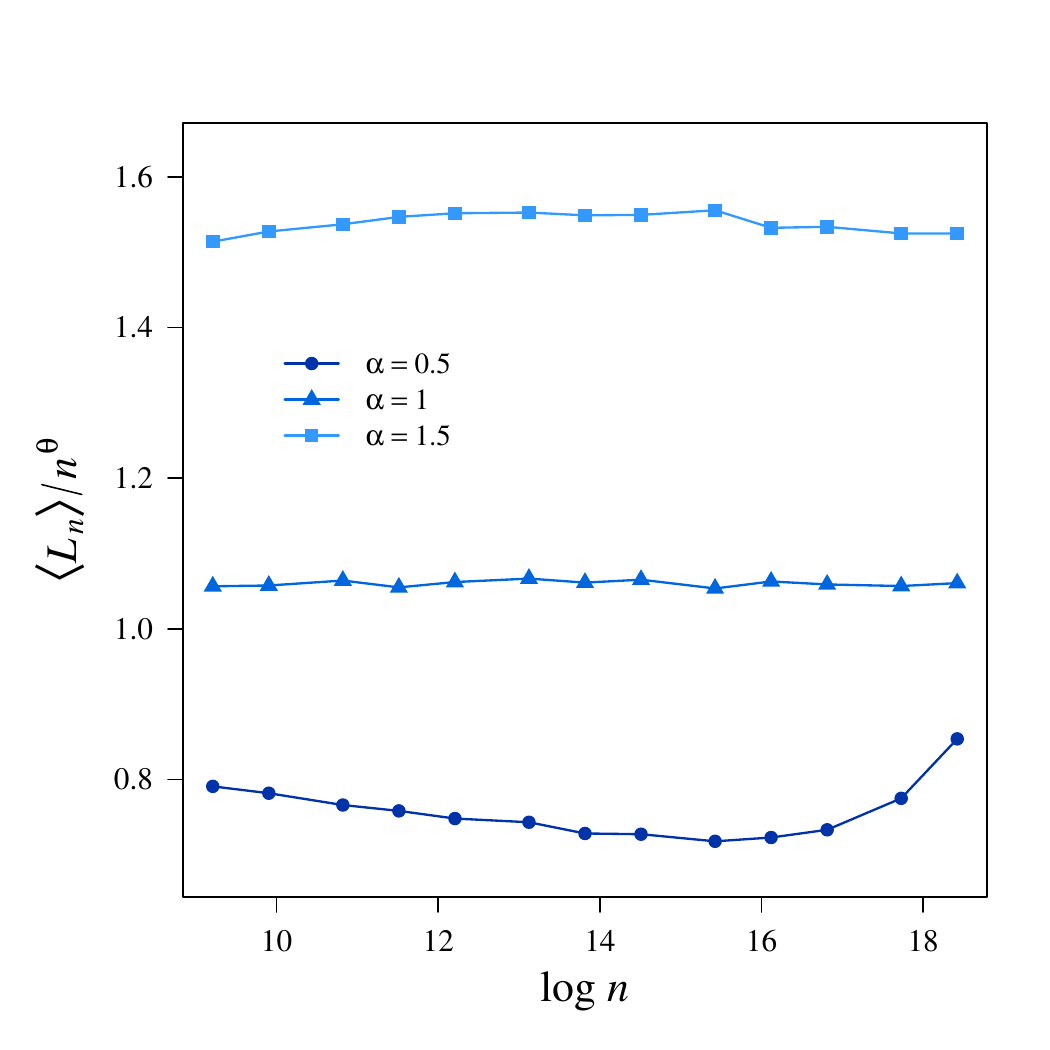}
\caption{Ratio $\langle{L_{n}}\rangle/n^{\theta}$ as a function of $\log{n}$ for $\alpha < 2$, where $\theta$ is the Model~II exponent from Table~\ref{tab:adjIIa} for each $\alpha$. For $\alpha=1$ and $\alpha=3/2$ the ratio is approximately constant, consistent with a pure power law. For $\alpha=1/2$ the ratio decreases up to $n \approx 10^{7}$ but then increases at the largest walk lengths.}
\label{fig:ratio_heavy}
\end{figure}

For $\alpha < 2$, the situation is different. Figure~\ref{fig:ratio_heavy} shows the ratio $\langle{L_{n}}\rangle/n^{\theta}$ as a function of $\log{n}$, using Model~II's fitted $\theta$ for each $\alpha$. For $\alpha=1$, the ratio is remarkably flat, confirming that the pure power law $\langle{L_{n}}\rangle \sim an^{\theta}$ with $\theta \approx 0.685$ provides an excellent description of the data. For $\alpha=3/2$, the ratio is similarly flat at $\langle{L_{n}}\rangle/n^{\theta} \approx 1.54$.

\begin{figure}[t]
\centering
\includegraphics[viewport= 10 10 480 460, scale=0.36, clip]{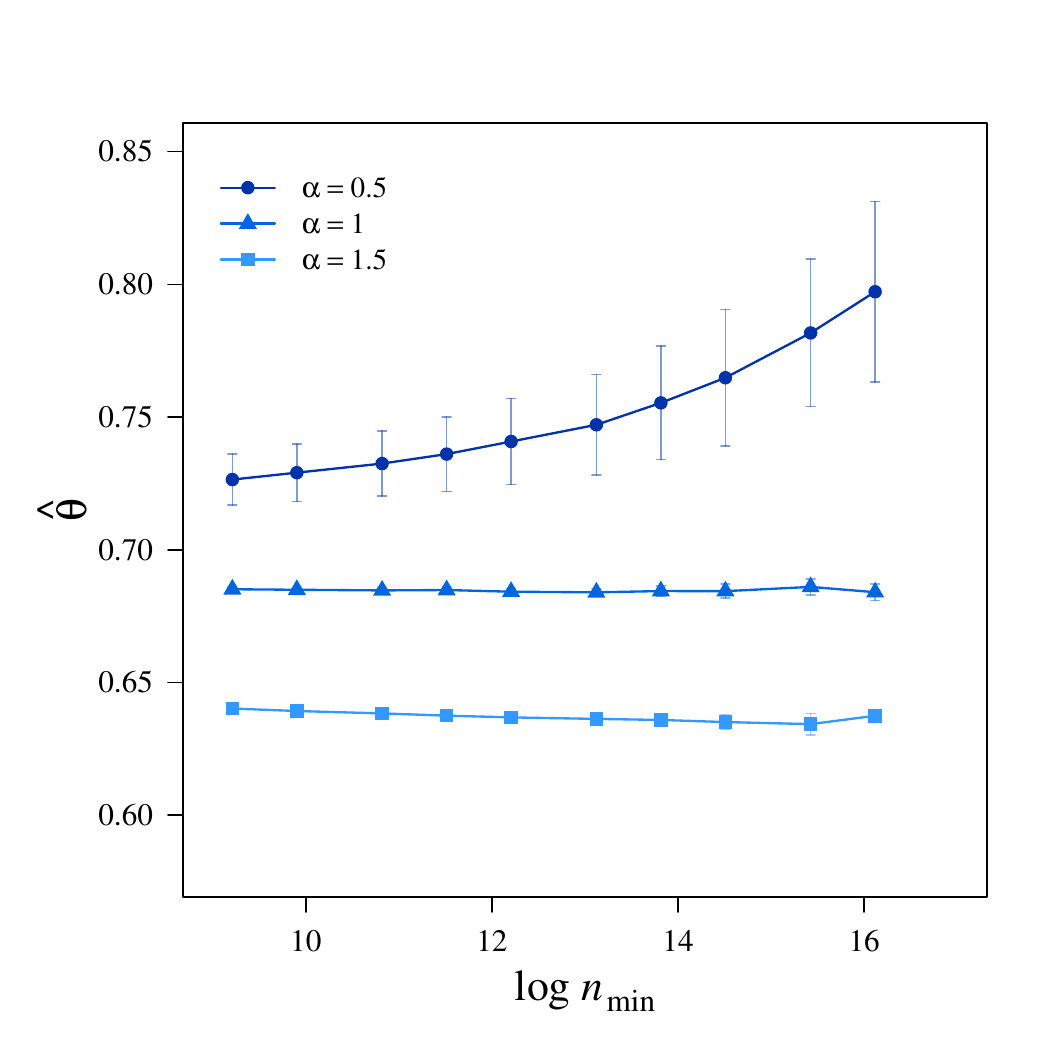}
\caption{Stability analysis for $\alpha < 2$. For $\alpha=1$, the estimated exponent $\hat{\theta}$ is stable at $\approx 0.685$ regardless of $n_{\min}$, confirming the pure power law. For $\alpha=3/2$, $\hat{\theta}$ is nearly stable with a very mild downward drift. For $\alpha=1/2$, $\hat{\theta}$ drifts upward with growing uncertainty, suggesting that the asymptotic regime has not been fully reached. Error bars indicate $\pm 2$ standard errors.}
\label{fig:stab_heavy}
\end{figure}

The stability analysis (Figure~\ref{fig:stab_heavy}) confirms these findings: $\hat{\theta}$ is essentially constant for $\alpha=1$ (at $\hat{\theta} \approx 0.685$) and displays only a very mild downward drift for $\alpha=3/2$ (from $0.640$ to $0.635$), consistent with a nearly pure power law with at most a minor logarithmic correction.

The case $\alpha=1/2$ displays anomalous behavior. Both the ratio plot and the stability analysis show that the behavior at the largest walk lengths ($n \gtrsim 10^{7}$) departs from the pattern observed at smaller~$n$: the ratio $\langle{L_{n}}\rangle/n^{\theta}$ turns upward, and the fitted $\hat{\theta}$ drifts from $0.73$ to $0.80$ as small-$n$ data is removed. The growing error bars reflect the reduced number of data points but also the increased variance of $L_{n}$ at large~$n$ for this very heavy tail, which was already noted in the exploratory analysis of the data. We interpret this as an indication that the asymptotic scaling regime has not been fully reached for $\alpha=1/2$ within the range of walk lengths studied, and that longer walks would be needed to settle the exponent. The estimate $\theta \approx 0.73$ obtained from fitting all data points should therefore be regarded with some caution. The slow convergence and increasing variability indicate that the effective exponent extracted from finite-size data may differ substantially from its eventual asymptotic value in this regime.

\subsubsection{Summary}

Table~\ref{tab:summary} summarizes our best estimates of the leading asymptotic behavior of $\langle{L_{n}}\rangle$ based on the analyses presented in this section. For $\alpha \leq 1$, the data are well described by a pure power law $\langle{L_{n}}\rangle \sim an^{\theta}$ with a varying exponent $\theta > 1/2$. For $\alpha \geq 2$ and the simple random walk, the data are consistent with $\theta = 1/2$ and a logarithmic correction, $\langle{L_{n}}\rangle \sim \sqrt{n}\mku(a+b\log{n})$. The behavior at $\alpha=3/2$ suggests that this value lies in a crossover region between the power-law regime ($\alpha < 2$) and the logarithmically corrected regime ($\alpha \geq 2$). In this intermediate regime, both mechanisms appear to contribute: while a pure power law provides a good effective description over the range of $n$ studied, the presence of a detectable logarithmic correction indicates that the asymptotic behavior may not yet be fully realized.

\begin{table}[h]
\caption{Summary of the best-supported asymptotic model for $\langle{L_{n}}\rangle$ as a function of the tail index $\alpha$.}
\centering
\begin{tabular}{c|ccc}
\hline
$\alpha$ & best model & leading behavior & $\theta$ \\ \hline
$1/2$    & II & $an^{\theta}$ & $\approx 0.73^{*}$ \\
$1$      & II & $an^{\theta}$ & $0.685(0)$ \\
$3/2$    & II & $an^{\theta}$ & $0.640(1)$ \\[2pt] 
\hline \\[-9pt]
$2$      & II & $\sqrt{n}\mku(a+b\log{n})$ & $1/2$ \\
$3$      & II & $\sqrt{n}\mku(a+b\log{n})$ & $1/2$ \\
$5$      & II & $\sqrt{n}\mku(a+b\log{n})$ & $1/2$ \\
$10$     & II & $\sqrt{n}\mku(a+b\log{n})$ & $1/2$ \\
SRW      & II & $\sqrt{n}\mku(a+b\log{n})$ & $1/2$ \\ \hline
\multicolumn{4}{l}{\footnotesize $^{*}$Subject to caution; see the discussion of $\alpha=1/2$ in Section~\ref{susub:caution}.}
\end{tabular}
\label{tab:summary}
\end{table}


\section{\label{sec:anova}Weighted nonlinear least squares and model comparison}

The analyses in Section~\ref{sec:lis} are based on sample means $\langle{L_{n}}\rangle$, which compress each Monte Carlo ensemble of $10{,}000$ values into a single data point per walk length. Working with the individual $L_{n}$ sample values rather than the group means allows us to estimate the variance at each walk length directly from the data, which enables proper inverse-variance weighting and valid $F$-tests for nested model comparison. We thus complement the previous analyses with a weighted nonlinear least squares (NLS) fit applied to the individual $L_{n}$ values, followed by an analysis of variance (ANOVA): given two candidate models in which one is a special case of the other (nested), the $F$-statistic measures whether the extra parameters of the larger model produce a reduction of the residual sum of squares beyond what would be expected by chance, with the corresponding $p$-value quantifying the evidence against the smaller (null) model. A general reference for the methods employed here is \cite{regression}. We note that the validity of the $F$-tests does not require individual $L_{n}$ values to be normally distributed; with $10{,}000$ replications per walk length, the central limit theorem ensures that the sample means are approximately normal, which is sufficient for the asymptotic theory underlying the weighted NLS inference.

We reparametrize the leading exponent as $\theta = 1/2+\delta$ and fit the full model
\begin{equation}
\label{eq:full}
L_{n} = n^{1/2+\delta}(a+b\log{n})
\end{equation}
to all $130{,}000$ observations ($10{,}000$ samples $\times$ $13$ walk lengths) for each value of $\alpha$, using the Levenberg-Marquardt algorithm as implemented in the R package \texttt{minpack.lm}. Because $\var(L_{n})$ grows with~$n$, we weight each of the $10{,}000$ observations by the inverse of the group variance $1/\var(L_{n}\,|\,\alpha,n)$, estimated from the sample.

For $\alpha < 2$, we test $H_{0}\mkern-2mu\colon b=0$ by comparing the full model (\ref{eq:full}) with the restricted model $L_{n} = an^{1/2+\delta}$, which corresponds to a pure power law. For $\alpha \geq 2$ and the simple random walk, we test $H_{0}\mkern-2mu\colon \delta=0$ by comparing (\ref{eq:full}) with the restricted model $L_{n} = \sqrt{n}\mku(a+b\log{n})$, which fixes $\theta=1/2$. This testing framework is summarized in Table~\ref{tab:tests}.

\begin{table}[t]
\caption{Summary of the nested model comparison. In both cases the restricted model is nested within the full model (\ref{eq:full}), and the comparison reduces to an $F$-test with $1$ and $129{,}997$ degrees of freedom.}
\centering
\begin{tabular}{c|ccc}
\hline
regime & $H_{0}$ & full model & restricted model \\ \hline
$\alpha < 2$ & $b=0$ & $n^{1/2+\delta}(a+b\log{n})$ & $an^{1/2+\delta}$ \\
$\alpha \geq 2$ & $\delta=0$ & $n^{1/2+\delta}(a+b\log{n})$ & $\sqrt{n}\mku(a+b\log{n})$ \\ 
\hline
\end{tabular}
\label{tab:tests}
\end{table}

Table~\ref{tab:anova} presents the results. For $\alpha=1/2$ and $\alpha=1$, the hypothesis $b=0$ is not rejected ($p=1.00$ and $p=0.17$, respectively), confirming that the pure power law is adequate. For $\alpha=3/2$, the hypothesis $b=0$ is rejected ($p<10^{-11}$), indicating that a logarithmic correction is already detectable at this value of $\alpha$, consistent with the transitional behavior noted in Section~\ref{sec:compare}. For all $\alpha \geq 2$ and the simple random walk, the hypothesis $\delta=0$ is also rejected (all $p < 10^{-46}$), with estimated values $\hat{\delta} \approx 0.024$--$0.035$.

\begin{table}[h]
\caption{Weighted NLS with ANOVA testing. For $\alpha < 2$, the test is $H_{0}\mkern-2mu\colon b=0$ (no logarithmic correction). For $\alpha \geq 2$, the test is $H_{0}\mkern-2mu\colon \delta=0$ ($\theta = 1/2$ exactly). With $\hat{\theta} = 1/2+\hat{\delta}$, the column $\hat{\delta}$ gives the estimate from the selected model. The $p$-values are obtained analytically from the upper tail of the $F$-distribution with $(1,129{,}997)$ degrees of freedom (R function \texttt{pf}); those smaller than $10^{-15}$ or so should be understood as ``essentially zero,'' the operative quantity being the magnitude of the $F$-statistic itself.}
\centering
\begin{tabular}{c|ccccc}
\hline
$\alpha$ & $H_{0}$ & $\hat{\delta}$ & SE & $F$ & $p$ \\ \hline
$1/2$ & $b=0$ & $0.221$ & $4\times 10^{-4}$ & $0.0$ & $1.00$ \\
$1$ & $b=0$ & $0.185$ & $4\times 10^{-4}$ & $1.9$ & $0.17$ \\
$3/2$ & $b=0$ & $0.185^{\dagger}$ & $0.003$ & $49.1$ & $2.4\times 10^{-12}$ \\[2pt] \hline \\[-9pt]
$2$ & $\delta=0$ & $0.024$ & $0.002$ & $208.0$ & $4.2\times 10^{-47}$ \\
$3$ & $\delta=0$ & $0.030$ & $0.002$ & $323.2$ & $3.6\times 10^{-72}$ \\
$5$ & $\delta=0$ & $0.029$ & $0.002$ & $357.0$ & $1.6\times 10^{-79}$ \\
$10$ & $\delta=0$ & $0.035$ & $0.002$ & $492.4$ & $6.9\times 10^{-109}$ \\
SRW & $\delta=0$ & $0.034$ & $0.002$ & $469.5$ & $6.3\times 10^{-104}$ \\
\hline
\multicolumn{6}{l}{\footnotesize $^{\dagger}$\,From the full model (\ref{eq:full}); $H_{0}$ rejected.}
\end{tabular}
\label{tab:anova}
\end{table}

The rejection of $\delta=0$ for $\alpha \geq 2$ does not contradict the conclusion from Section~\ref{sec:compare} that $\theta=1/2$ is the correct asymptotic exponent. The ANOVA tests whether the \textit{finite-sample} data are better described by $\theta = 1/2+\delta$ than by $\theta=1/2$ exactly, and with $130{,}000$ observations the test has very high power to detect even minute departures from the null. The estimated $\hat{\delta} \approx 0.03$ is precisely the finite-size bias that the stability analysis (Figure~\ref{fig:stab_finite}) reveals graphically: when a pure power law is fitted to finite data that follow $\sqrt{n}\mku(a+b\log{n})$, the effective exponent is necessarily elevated above $1/2$, with the discrepancy shrinking logarithmically as longer walks are included. The ANOVA thus quantifies the finite-size artifact rather than identifying a genuine departure from $\theta=1/2$. By contrast, the rejection of $b=0$ at $\alpha=3/2$ is corroborated by the mild downward drift in the stability analysis (Figure~\ref{fig:stab_heavy}), which indicates that the logarithmic correction is a genuine feature of the data rather than an artifact of finite sample size.


\section{\label{sec:dist}Distributional diagnostics}

The analyses in the preceding sections focus on the scaling of $\langle{L_{n}}\rangle$ with~$n$. We now turn to the shape of the distribution of $L_{n}$ itself. If $\log{L_{n}}$ is approximately normal for each $(\alpha,n)$, then $L_{n}$ is approximately lognormal, a property that characterizes the fluctuations of the weak LIS and has implications for the choice of regression model. In particular, the approximate homoscedasticity of $\log{L_{n}}$ for $\alpha \geq 1$ validates the use of unweighted regression on the log scale in Section~\ref{sec:adjII}, while the growing variance of $L_{n}$ itself on the original scale motivates the inverse-variance weighting adopted in Section~\ref{sec:anova}. 

\begin{figure}[t]
\centering
\includegraphics[viewport= 0 0 600 240, scale=0.60, clip]{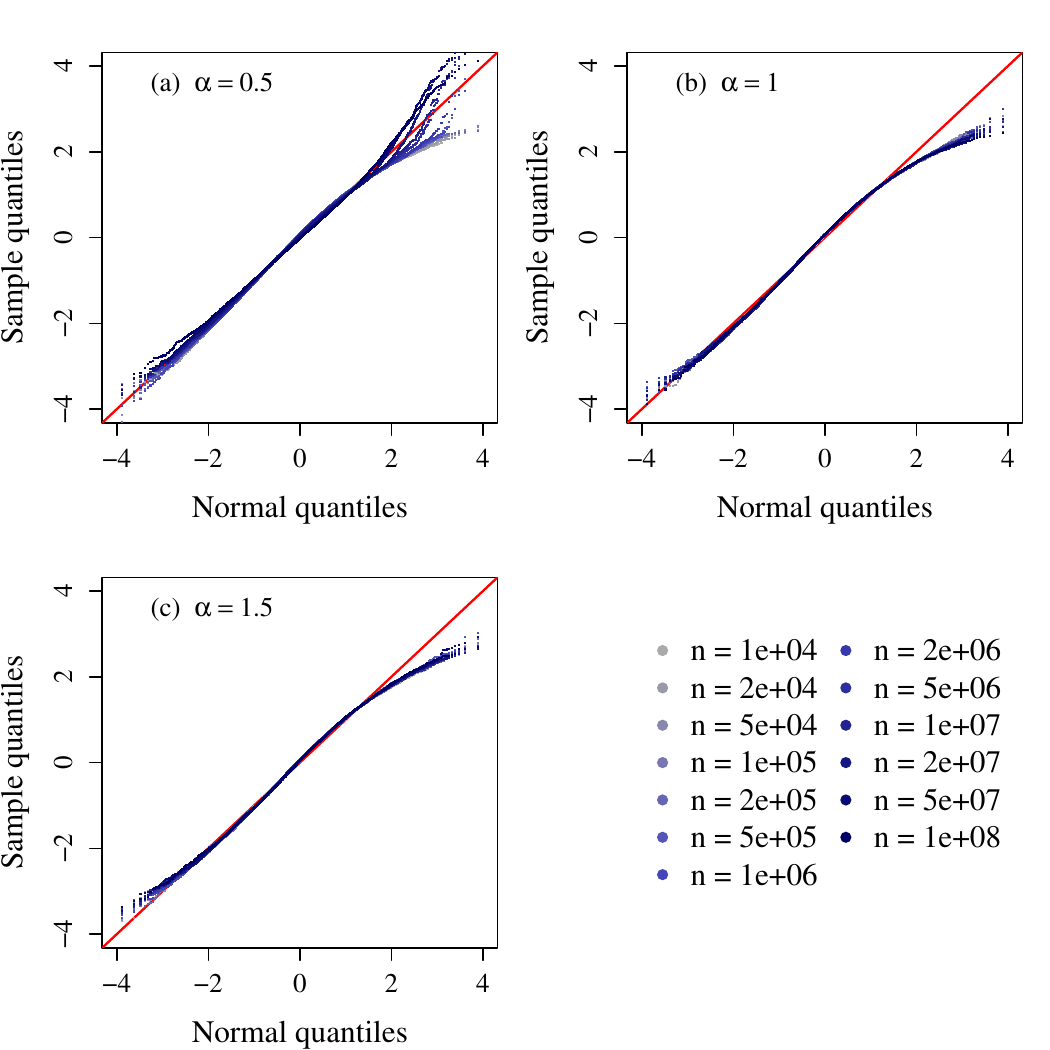}
\caption{Normal Q-Q plots of standardized $\log{L_{n}}$ for $\alpha < 2$. Each panel overlays the $13$ walk lengths, with darker shades corresponding to larger~$n$. The red line marks the identity. The points track the normal closely in the bulk, with light-tailed departures at the extremes.}
\label{fig:qq_heavy}
\end{figure}

\begin{figure}[t]
\centering
\includegraphics[viewport= 0 0 600 490, scale=0.60, clip]{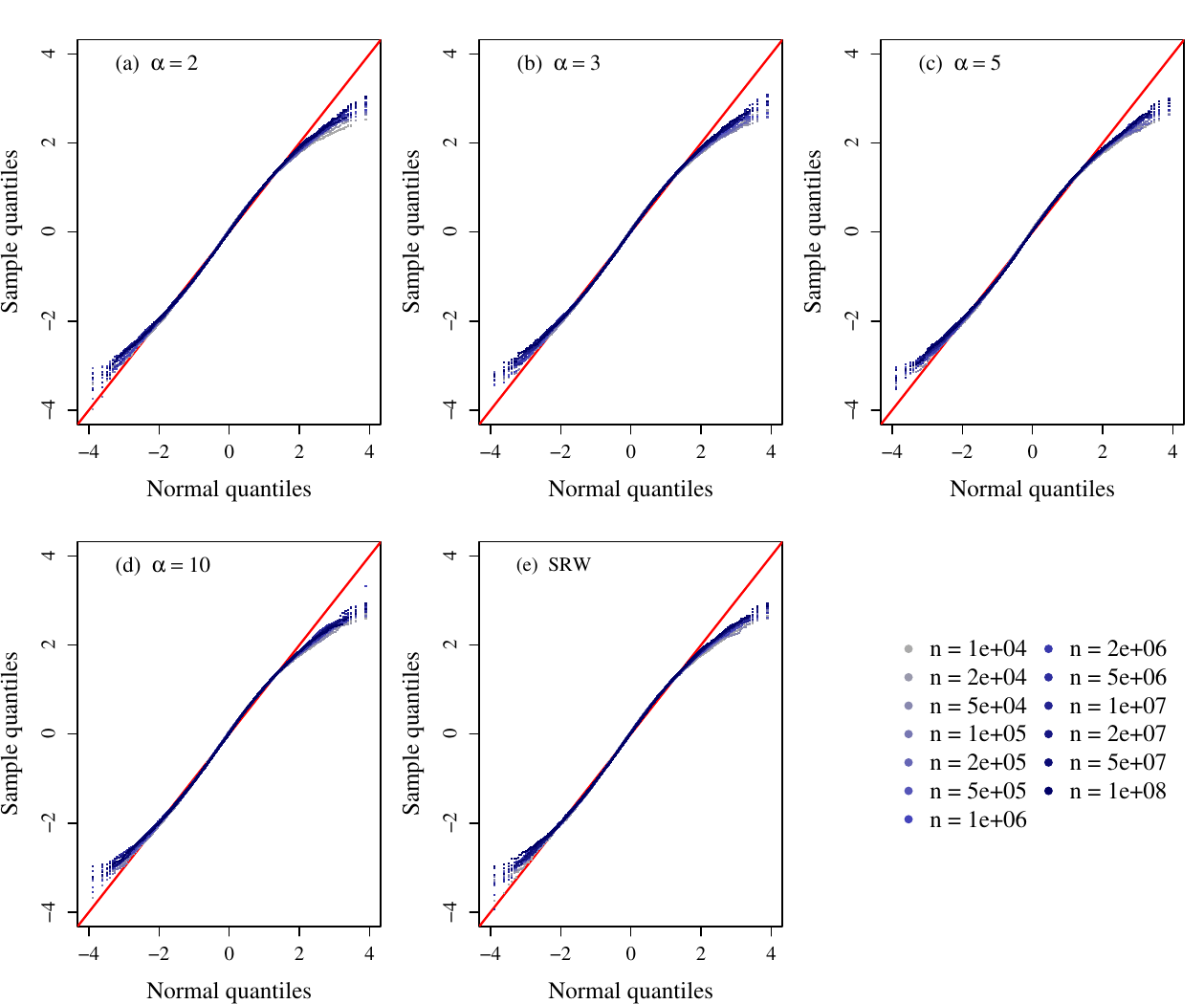}
\caption{Normal Q-Q plots of standardized $\log{L_{n}}$ for $\alpha \geq 2$ and the simple random walk. In each panel, the $13$ walk lengths from $n = 10^{4}$ to $n = 10^{8}$ are overlaid, with darker shades of blue corresponding to larger~$n$. The pattern of light-tailed departures is consistent across all values of $\alpha$.}
\label{fig:qq_finite}
\end{figure}

Figure~\ref{fig:qq_heavy} shows normal Q-Q plots of the standardized $\log{L_{n}}$ for $\alpha < 2$, with all $13$ walk lengths overlaid in each panel, and Figure~\ref{fig:qq_finite} shows the corresponding plots for $\alpha \geq 2$ and the simple random walk. In each case, the $10{,}000$ values of $\log{L_{n}}$ at a given~$n$ are standardized to zero mean and unit variance and plotted against the theoretical normal quantiles. The points lie close to the identity line across the bulk of the distribution for all $\alpha$ and~$n$, with deviations confined to the extreme tails ($\abs{z} \gtrsim 3$).

The departures behave in a roughly systematic manner instead of an erratic one: the sample quantiles in the tails fall slightly below the normal reference line, indicating that the tails of $\log{L_{n}}$ are systematically lighter than Gaussian. This pattern is consistent across all $\alpha \geq 1$ and~$n$ and becomes more pronounced as~$n$ grows. For $\alpha = 1/2$, the tails are roughly Gaussian at moderate~$n$ but develop positive skewness at the largest walk lengths ($n \geq 5 \times 10^{7}$), consistent with the anomalous behavior already noted in Section~\ref{sec:compare}.

\begin{figure}[t]
\centering
\includegraphics[viewport= 20 10 600 460, scale=0.45, clip]{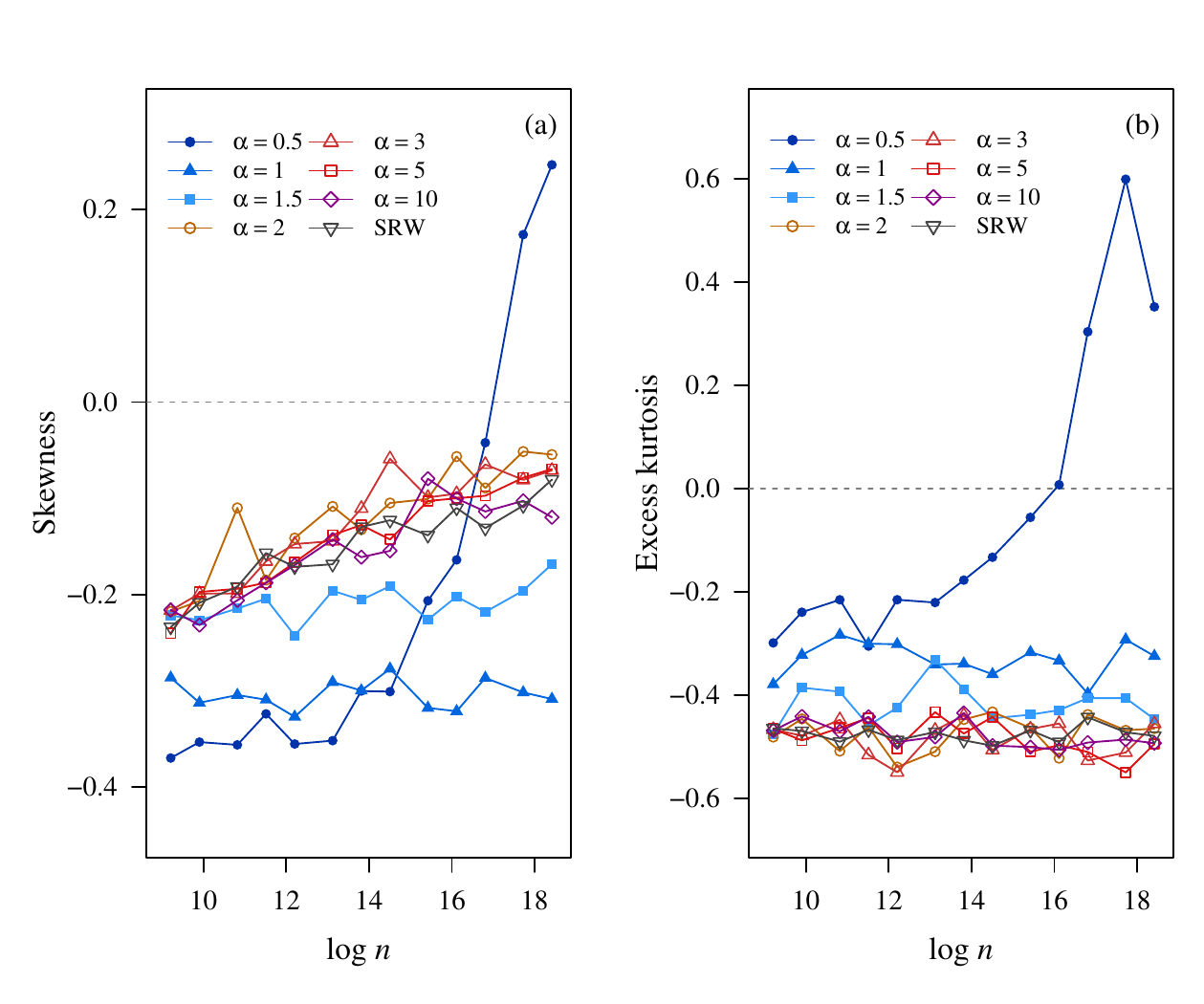}
\caption{Sample skewness~(a) and excess kurtosis~(b) of $\log{L_{n}}$ as a function of $\log{n}$ for all~$\alpha$. The dashed line marks the normal value of zero. The distributions are mildly left-skewed and platykurtic (lighter tails than normal) for all $\alpha \geq 1$.}
\label{fig:skku}
\end{figure}

To quantify the departures, Figure~\ref{fig:skku} shows the sample skewness and excess kurtosis of $\log{L_{n}}$ as a function of $\log{n}$ for all~$\alpha$. The skewness is mildly negative (typically $-0.35$ to $-0.05$), indicating a slight left skew, and tends toward zero as~$n$ increases for $\alpha \geq 2$. The excess kurtosis is negative throughout (typically $-0.55$ to $-0.30$), confirming the lighter-than-normal tails seen in the Q-Q~plots. The exception is again $\alpha=1/2$, where both the skewness and kurtosis drift toward positive values at large~$n$. The magnitudes are modest: $\abs{\text{skewness}} < 0.37$ and $\abs{\text{excess kurtosis}} < 0.55$ for all $\alpha \geq 1$, well within the range compatible with approximate normality for practical purposes, usually taken to be $\abs{\text{skewness}} < 0.5$ and $\abs{\text{excess kurtosis}} < 1$.

We also examined the variance of $\log{L_{n}}$ across~$n$ within each~$\alpha$. For $\alpha \geq 1$, the variance is approximately constant in~$n$ ($\var(\log{L_{n}}) \approx 0.14$--$0.17$), consistent with a homoscedastic normal model on the log scale. For $\alpha = 1/2$, the variance grows from $0.13$ to $0.21$ over the range of~$n$ studied, reflecting the increased dispersion at the largest walk lengths. In terms of the original (untransformed) variable, the corresponding coefficient of variation across the $10{,}000$ Monte Carlo realizations, $\sigma(L_{n})/\langle L_{n}\rangle = \sqrt{e^{\var(\log L_{n})}-1}$ under the lognormal approximation, takes values $\approx 0.39$--$0.43$ for $\alpha \geq 1$ at all walk lengths, growing to $\approx 0.48$ for $\alpha = 1/2$ at $n = 10^{8}$. Thus, although the relative dispersion of $L_{n}$ across realizations does increase as the tail becomes heavier, it remains moderate (under one half of the mean) across the entire range of parameters studied.

\begin{figure}[h]
\centering
\includegraphics[viewport= 0 0 600 240, scale=0.60, clip]{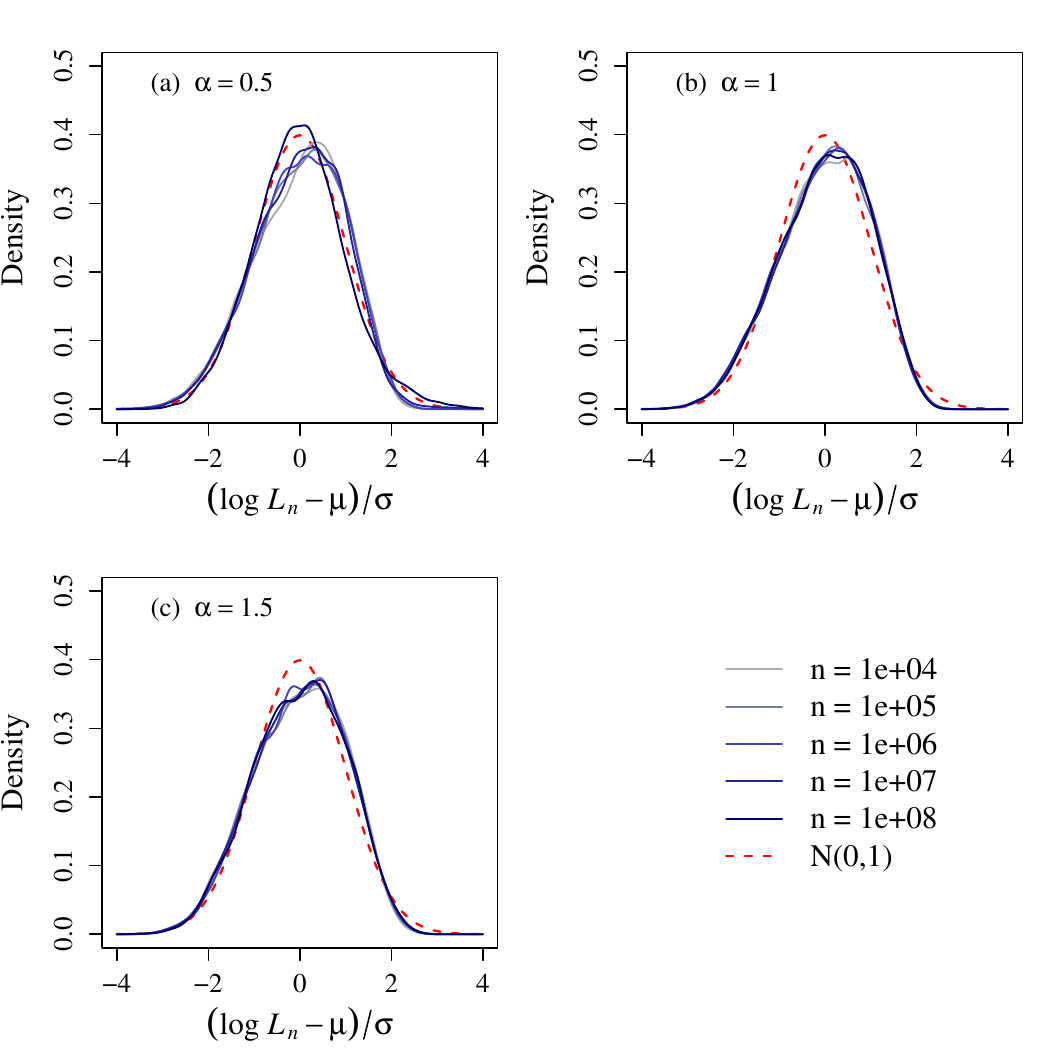}
\caption{Kernel density estimates of standardized $\log{L_{n}}$ for $\alpha < 2$, at five selected walk lengths ($n = 10^{4}$, $10^{5}$, $10^{6}$, $10^{7}$, $10^{8}$). The dashed red curve is the standard normal density.}
\label{fig:dens_heavy}
\end{figure}

\begin{figure}[t]
\centering
\includegraphics[viewport= 0 0 600 490, scale=0.60, clip]{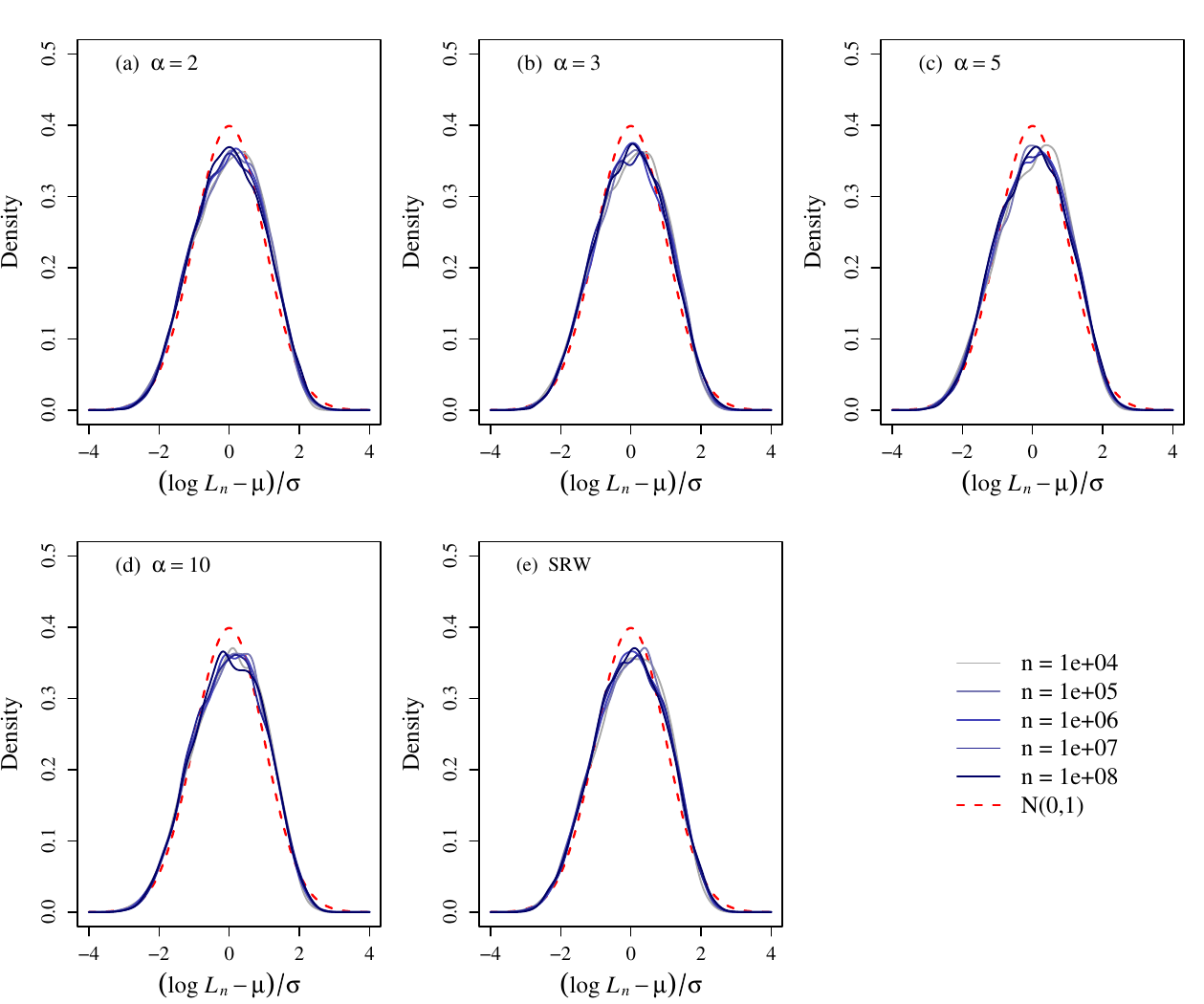}
\caption{Kernel density estimates of standardized $\log{L_{n}}$ for $\alpha \geq 2$ and the simple random walk, at the same five walk lengths as in Figure~\ref{fig:dens_heavy}. The dashed red curve is the standard normal density.}
\label{fig:dens_finite}
\end{figure}

To visualize the distributional shape directly, Figure~\ref{fig:dens_heavy} shows kernel density estimates of standardized $\log{L_{n}}$ for $\alpha < 2$ and Figure~\ref{fig:dens_finite} shows the same for $\alpha \geq 2$ and the simple random walk, each overlaid with the standard normal density. In every panel, the empirical densities track the Gaussian curve closely, with the slight platykurtic character (a peak marginally above and tails marginally below the normal) visible at larger~$n$, consistent with the negative excess kurtosis reported above. This platykurtic character of $\log L_n$ indicates that the distribution decays faster than a true lognormal in the extremes. 

\begin{figure}[t]
\centering
\includegraphics[viewport= 0 0 600 490, scale=0.55, clip]{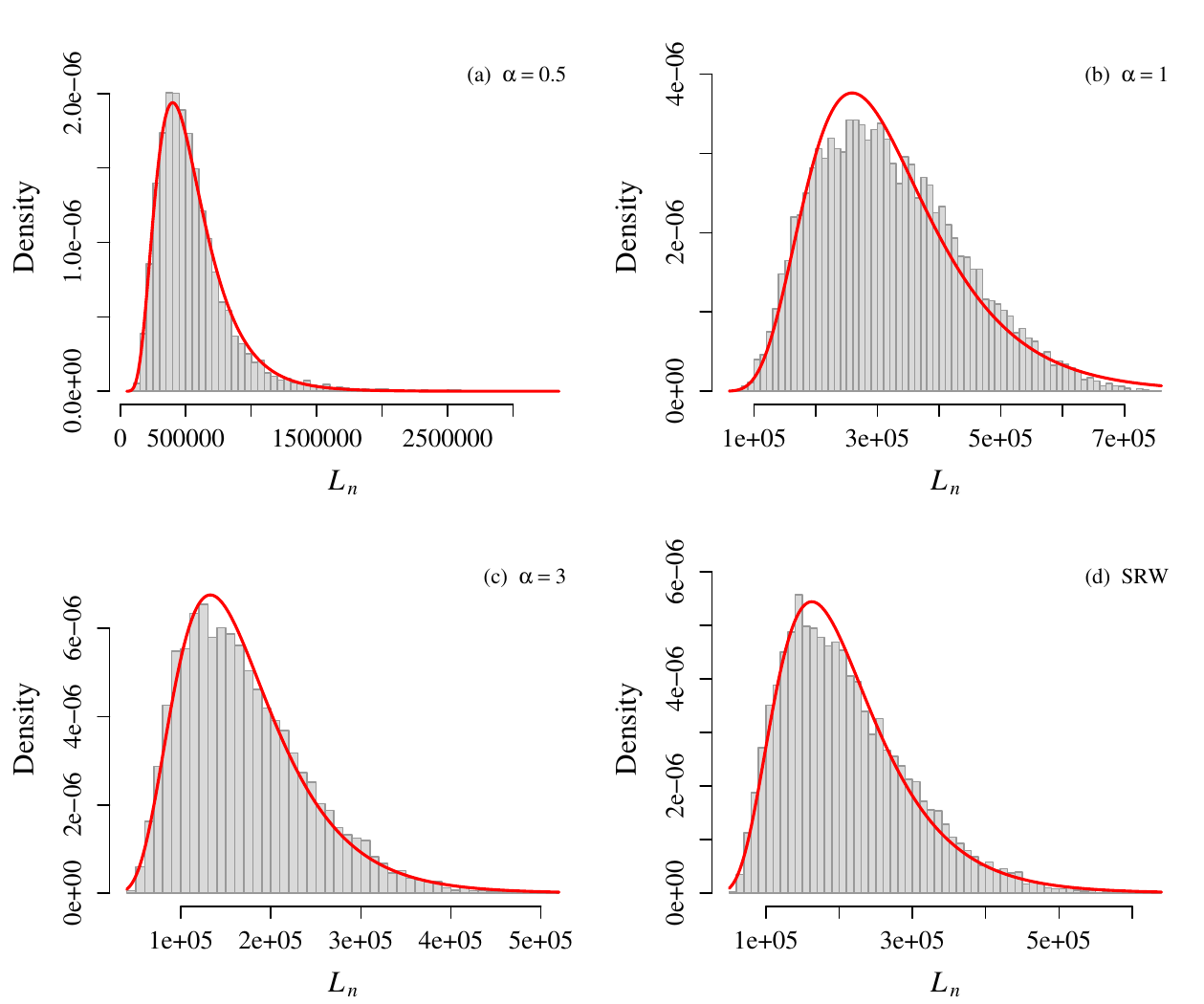}
\caption{Histograms of $L_{n}$ ($n = 10^{8}$ samples) for four representative values spanning both the heavy-tailed and finite-variance regimes, including the simple random walk, with the fitted lognormal density (solid red curve).}
\label{fig:hist_lognormal}
\end{figure}

Figure~\ref{fig:hist_lognormal} complements the density plots by showing histograms of the raw $L_{n}$ at $n = 10^{8}$ for four representative values of~$\alpha$, together with the lognormal density fitted by matching the sample mean and variance of $\log{L_{n}}$. The fitted curves closely follow the observed histograms, confirming that the lognormal model captures both the location and the characteristic right-skewed shape of the $L_{n}$ distribution.

We do not report formal goodness-of-fit tests because with $10{,}000$ observations per group, even the small systematic departures documented above would lead to rejection; the graphical diagnostics are more informative about the practical adequacy of the lognormal approximation. These diagnostics support the conclusion that $L_{n}$ is approximately lognormally distributed for all~$\alpha$ and~$n$ studied, with the approximation being best for $\alpha \geq 1$ and deteriorating for $\alpha = 1/2$ at the largest walk lengths.


\section{\label{sec:discuss}Discussion}

We have provided a characterization of the scaling behavior of the weak LIS of discrete heavy tailed random walks on the integers together with that of the limiting case, as $\alpha \to \infty$, of a simple random walk. The study extends previous work \cite{lisjpa,hartmann,lispre,ultrafat} by providing data and analyses on the LIS of random walks with discrete heavy tailed distribution of increments. We developed a systematic model-selection apparatus (effective exponent plots, ratio plots, stability analysis, and weighted nonlinear least squares with ANOVA model comparison) that goes beyond the exploratory fits of \cite{lisjpa,hartmann,lispre,ultrafat} and allows one to distinguish between the scaling forms $\sqrt{n}\mku\log{n}$ and $n^{\theta}$.

For the random walks with heavy-tailed increments, \ie, with $\alpha<2$, the LIS investigated here displayed the same qualitative behavior found before for the LIS of random walks with continuous heavy-tailed distribution of increments. In particular, all exponents $\theta$ fall within the rigorous bounds (\ref{eq:infinite}). Quantitatively, however, the values of $\theta$ obtained here for the discrete Zipf-distributed walks differ from those reported for continuous heavy-tailed walks at the same nominal $\alpha$ \cite{lisjpa,lispre,ultrafat} by amounts that exceed the quoted statistical uncertainties. This is to be expected: the underlying increment distributions are different (discrete Zipf vs.\ continuous Pareto, Student-$t$, or stable laws), the discrete setting permits plateaus that contribute non-trivially to the weak LIS in a way that continuous walks cannot replicate, and the model-selection methodology adopted here (NLS with inverse-variance weighting and ANOVA) differs from the exploratory fits used in earlier studies. The agreement with prior work is therefore at the level of the universal qualitative pattern --- a continuously varying exponent $\theta(\alpha)>1/2$ for $\alpha<2$ saturating to $\theta=1/2$ for $\alpha\geq 2$ --- rather than at the level of individual exponent values, which are non-universal across distributions and walk types. The ratio plots and stability analysis presented in Section~\ref{sec:compare} support a pure power law $\langle{L_{n}}\rangle \sim an^{\theta}$ for $\alpha \leq 1$, with a continuously varying exponent $\theta$ that increases as $\alpha$ decreases. The case $\alpha=3/2$ appears transitional, and the case $\alpha=1/2$ requires caution due to anomalous behavior at the largest walk lengths studied. The theoretical upper bound for $\theta$, obtained by considering an ultra-fat tailed random walk without any finite moment, is $\theta < 0.815$ \cite{pemantle,ultrafat}. We note that the ANOVA evidence at $\alpha=3/2$, together with the gradual evolution of the effective exponents, suggests that the crossover between the power-law and logarithmic-correction regimes is not sharp in finite-size data, even though the theoretical boundary is at $\alpha=2$ (finite vs.\ infinite variance). The apparent onset of pure power-law scaling below $\alpha=2$ may thus reflect a slow crossover rather than a distinct phase boundary. Moreover, the rate at which the asymptotic scaling regime is reached varies considerably with $\alpha$: for $\alpha \leq 1$, the heavy-tail-dominated power-law behavior sets in relatively quickly, whereas for $\alpha=1/2$ the extreme volatility of the walk demands longer sequences, and for $\alpha \geq 2$ convergence to the $\sqrt{n}\mku\log{n}$ form is inherently slow due to the logarithmic correction itself.

For $\alpha \geq 2$ and the simple random walk, the data are consistent with $\theta = 1/2$ and a logarithmic correction, $\langle{L_{n}}\rangle \sim \sqrt{n}\mku(a+b\log{n})$. The stability analysis shows that the elevated exponent $\theta \simeq 0.54$ found in Model~I is an artifact of the $\log{n}$ term being absorbed into the power law. The weighted NLS analysis of Section~\ref{sec:anova} confirms this interpretation: although the ANOVA formally rejects $\delta=0$ ($\theta=1/2$), the estimated $\hat{\delta} \approx 0.03$ is precisely the finite-size bias expected from the logarithmically slow convergence. The coefficients $a$ and $b$ found here differ significantly from those obtained for random walks with continuous finite-variance distributions of increments ($a \approx b \approx 0.36$) \cite{lisjpa,hartmann,lispre}, reflecting the enhanced role of the logarithmic correction in the discrete setting, where it is rigorously established \cite{angel}. 

The presence of the logarithmic correction in our discrete walks, with coefficients substantially larger than those found in the continuous case, suggests that the $\log{n}$ correction is intrinsic to the scaling, becoming more prominent when the distribution of increments is itself discrete. The unexpected appearance of this logarithmic correction in previous numerical studies of continuous walks may be influenced by the intrinsic discreteness of floating-point representations in the simulations (double-precision floating-point arithmetic imposes an effective lattice spacing of order $10^{-16}$, which could in principle allow plateau-like effects to emerge at very large~$n$). The study of the LIS of random walks with uniformly distributed increments $\pm 1, \dots, \pm k$, $k>1$ fixed, could shed further light on this point, since one would then observe how the coefficients $a$ and $b$ evolve with $k$ as the distribution interpolates between the discrete (small $k$) and continuous ($k \to \infty$) regimes; the data at both endpoints are already available.

The distributional diagnostics of Section~\ref{sec:dist} indicate that $L_{n}$ is well approximated by a lognormal distribution for all $\alpha$ and~$n$ studied, with the approximation being best for $\alpha \geq 1$. The mild negative skewness and negative excess kurtosis of $\log{L_{n}}$ suggest that the tails of the distribution are lighter than Gaussian, rather than heavier. The lognormality of $L_{n}$ is an empirical observation for which we do not yet have a theoretical explanation. Lognormal distributions typically arise from multiplicative growth mechanisms or from sums of correlated contributions in logarithmic variables. The LIS is a global combinatorial functional of the sequence that cannot be decomposed into many roughly independent contributions along the random walk in any obvious sense. Whether this approximate lognormality extends to other families of random walks, whether the parameters of the lognormal depend on~$\alpha$ in a predictable way, and, ultimately, whether it can be derived rigorously from the combinatorial structure of the LIS (e.\,g., by identifying a multiplicative structure in the accumulation of admissible subsequences or through a decomposition of the patience sorting tableau) remain open questions.


\section*{\label{sec:credit}CRediT authorship contribution statement}

J. R. G. Mendonça: Conceptualization, Funding acquisition, Methodology, Software, Visualization, Writing -- original draft, Writing -- review \&\ editing.

M. V. Freire: Methodology, Software, Visualization, Writing -- original draft, Writing -- review \&\ editing.


\section*{\label{sec:interests}Declaration of competing interest}

The authors declare that they have no known competing financial interests or personal relationships that could have appeared to influence the work reported in this paper.


\section*{\label{sec:ack}Acknowledgments}

JRGM thanks FAPESP, the São Paulo State Research Foundation, Brazil, for partial support under grant no.~\mbox{2020/04475-7}.


\end{document}